\documentclass[notitlepage]{article}

\usepackage[a4paper,top=1.5cm,bottom=1.5cm,left=1.5cm,right=1.5cm,marginparwidth=1.75cm]{geometry}

\usepackage{placeins}
\usepackage{lineno,hyperref}
\usepackage[numbers]{natbib}
\usepackage{graphicx}
\usepackage{subfig}
\usepackage{multirow}
\usepackage{amsmath}
\usepackage{algorithm}
\usepackage{algpseudocode}
\usepackage{makecell} 
\usepackage{amsmath}
\usepackage{dblfloatfix}
\usepackage{amssymb}
\usepackage{booktabs}
\usepackage{xcolor}
\usepackage{mathtools}
\usepackage{natbib}
\usepackage{xfrac}
\usepackage[symbol]{footmisc}
\let\printorcid\relax 
\definecolor{ForestGreen}{RGB}{34,139,34}
\definecolor{BrickRed}{RGB}{203,65,84}
\usepackage{titling}
\usepackage{authblk}
\usepackage{caption} 
\title{A copula-based boosting model for time-to-event prediction with dependent censoring}    
\date{\vspace{-5ex}}

\begin{document}

\author[1,*]{Alise Danielle Midtfjord}
\author[1]{Riccardo De Bin}
\author[1]{Arne Bang Huseby}
\affil[1]{Department of Mathematics, University of Oslo, Norway}

\let\WriteBookmarks\relax
\def\floatpagepagefraction{1}
\def\textpagefraction{.001}

\twocolumn[
  \begin{@twocolumnfalse}
    
    \maketitle

\begin{abstract}
A characteristic feature of time-to-event data analysis is possible censoring of the event time. Most of the statistical learning methods for handling censored data are limited by the assumption of independent censoring, even if this can lead to biased predictions when the assumption does not hold. This paper introduces Clayton-boost, a boosting approach built upon the accelerated failure time model, which uses a Clayton copula to handle the dependency between the event and censoring distributions. By taking advantage of a copula, the independent censoring assumption is not needed any more. During comparisons with commonly used methods, Clayton-boost shows a strong ability to remove prediction bias at the presence of dependent censoring and outperforms the comparing methods either if the dependency strength or percentage censoring are considerable. The encouraging performance of Clayton-boost shows that there is indeed reasons to be critical about the independent censoring assumption, and that real-world data could highly benefit from modelling the potential dependency. 
\end{abstract}

\vspace{1cm}

  \end{@twocolumnfalse}]
  
\section{Introduction}
The analysis of time-to-event data is an important topic in statistics. It deals with response variables that record the time until a specific event happens, for example death due to a specific disease or failure of a mechanical component. While very common in biomedicine \citep{khan2016variable, negassa2005tree, dang2021penalized, porzelius2010sparse} and engineering \citep{louit2009practical, mcpherson2018reliability, carvalho2019systematic}, where it takes the name of survival analysis and reliability theory, respectively, the analysis of time-to-event data is used in many other fields as well, such as economics \citep{OrbeAl2001}, ecology \citep{roksvaag2022probabilistic}, and agronomy \citep{OnofriAl2018}. 

A characterizing feature of time-to-event data  is the presence of censored observations, i.e., some statistical units are not observed until the event of interest appears, but up to (or from) a different time-point. Examples where we only observe a lower bound of the time are when patients exit a clinical study before achieved test result or when substituting a machinery component before its failure \citep{Lagakos1979}. An example where we only observe an upper bound of the time, or observe an interval which the times lie within, is when systems are investigated at a specific time point, and components which have failed are identified at that time \citep{samuelsen1994interval,skutlaberg2018partial,gaasemyr2001bayesian}. Specific methods have been developed to handle data involving censored observations, including the proportional hazard model \citep{Cox1972}, the first hitting threshold models \citep{EatonWhitmore1977} and the accelerated failure time models \citep{Pike1966}.

The development of advanced machine learning in the recent years has also influenced the analysis of time-to-event data. Techniques as neural networks (e.g., \cite{KatzmanAl2018,KvammeAl2019}), support vector machine \citep{VanAl2007,KhanZubek2008}, random forests \citep{IshwaranAl2008,HornungWright2019}, and many other (for a recent review, see\cite{WangAl2019} have been adapted to work in this context. In this paper, we will focus on boosting techniques, which have also been successfully implemented for modelling time-to-event data \citep{MayrAl2014}. For example, boosting counterparts of the proportional hazard model (e.q., \cite{BinderSchumacher2008}), first hitting time model \citep{DebinStikbakke2022} and accelerated failure time model \citep{SchmidHothorn2008} are available in the literature, and well implemented in statistical software \citep{Debin2016}.
 
As many of the classical models they are based on, however, currently available boosting algorithms are limited by the assumption of independent censoring. This means that they require the censorship mechanism to be independent of the event of interest. While plausible in some cases, this assumption may not hold in many other \citep{CzadoVankeilegom2021}. Typical examples are patients that drop out from a study due to some reasons related to the therapy or by the effect of a competing risk \citep{Lagakos1979}. The consequences of incorrectly assuming independent censoring range from the wrong assessment of the effect of a variable to the overestimation/underestimation of the survival probabilities, and several research papers have clearly stated that the independent censoring assumption deserves special scrutiny \citep{kaplan1958nonparametric,leung1997censoring,EmuraChen2018}. Illustrative examples can be found in \cite{HuangZhang2008}. Their simulation study shows how much one can overestimate the survival probabilities when incorrectly assuming independent censoring (Figure 1(a) in \cite{HuangZhang2008}) and how much bias can be introduced in the estimate of the regression coefficients (Table 1 in \cite{HuangZhang2008}). Their application also shows how much the p-value related to a covariate can vary when the model is fitted assuming or not assuming independence censoring, at the point that a risk factor can appear statistically significant when it is not (or the other way around). Many other simulation (see, e.q., \cite{wilhems,campigotto2014impact}) or real data (see, e.g.,\cite{Liu2017}) examples can be found in the literature on this issue.

While there is much work in the classical statistical framework on this topic (see, among many other, \cite{Link1989,HuangWolfe2002,ScharfsteinRobins2002,Siannis2004,LuZhang2012,JacksonAl2014,DettoniAl2020}), less attention has been devoted to extend statistical learning methods to the case of dependent censoring. A notable exception is \cite{MoradianAl2019}, where the survival version of random forests is adapted to tackle the issue. The goal of our paper is to help filling this gap. In particular, we introduce a gradient boosting model that handles time-to-event data with dependent censoring. First developed in the machine learning community \citep{Schapire1990,Freund1995} and then translated into the statistical world \citep{FriedmanAl2000}, gradient boosting is one of the most effective machine learning tools currently available \citep{MayrAl2014b}. Especially, recent implementations like XGBoost \citep{ChenGuestrin2016}, LightGBM \citep{KeAl2017} and CatBoost \citep{DorogushAl2018} have been proved to work extremely well in many situations \citep{midtfjord2021,nielsen2016tree}, and a boosting model which deals with dependent censoring is highly desirable.

We will tackle the problem with the help of copulas \citep{Sklar1959}. Copulas are cumulative distribution functions with uniform marginals, which can be interpreted as the dependence structure between two or more  distributions. In this paper, we use a Clayton copula when modelling the dependency between the event time and the censoring time. This approach is not new in the literature, and started with the seminal work of \cite{ZhengKlein1995}. Here, a generalization of the Kaplan-Meier estimator, called copula-graphic, was developed to handle dependent censoring. Extensions of that work include \cite{RivestWells2001}, which focus on Archimedean copulas, and \cite{BraekersVeraverbeke2005}, which study it in the fixed-design situation. Among more recent works, we mention \cite{DeunaalvarezVeraverbeke2017}, that consider the left-truncation case, and \cite{CzadoVankeilegom2021}, that relax the assumption that the parameter defining the copula function is known. For a full description of the use of copula to model dependent censoring in time-to-event data problems, we refer the reader to the book of \cite{EmuraChen2018}.

The rest of the paper is organized as follow. In Section \ref{methods} we describe dependent censoring and introduce the basic concept of our approach, the Clayton Copula, the Accelerated Failure time Model, and the boosting algorithm. The novel approach is presented in Section \ref{sec:copulaXBoost} and evaluated via simulation and on real data in Section \ref{applications}. Section \ref{conclusions} ends the paper with some remarks.

section{Methods}\label{methods}

\subsection{Time-to-event prediction and \newline likelihood-based inference}
In survival analysis, the term survival time, or \textit{event time}, refers to the time progressed from an origin to the occurrence of an event. While the target variable is commonly referred to as "time", it can also consist of other units such as cycles, rounds, or even friction, as will be seen in Section \ref{applications}. One common factor for the different applications of time-to-event prediction is the presence of censored data. There are three main ways of censoring: In right censoring, the event time is known to be higher than a certain value, while in left censoring, the event time is known to be lower than a certain value. In interval censoring, the event time is known to be between two values. This paper will focus on right censoring, as this is the most common way of censoring. However, the proposed methodology is easily extended to handle left- and interval censoring by making a smaller change in the likelihood function. 

In addition to the tree ways of censoring, there are some different characteristics with the censoring. A \textit{Type I} censoring means that the event is censored only if it happens after a pre-specified time. This is also called \textit{administrative censoring}, and all remaining subjects at the specified time are right censored. A \textit{Type II} censoring means that an experiment stops after a pre-specified number of events has happened, and the rest of the subjects are right censored. When the censoring is \textit{Independent and non-informative}, every subject has a probability of being censored which is statistically independent of the event time. On the contrary, if a censoring is referred to as \textit{Dependent}, the subject is censored by a mechanism related to the event time. This paper concerns the latter censoring characteristic and proposes a method for making statistical learning methods useful for this type of censoring, and not only in the case of independent and non-informative censoring, which is often assumed in the literature. Before going into the details, we review the classical terminology of time-to-event data, which is further used in this paper.

Consider two random variables: $T$ is the event time and $U$ is the censoring time. The two variables are mutually exclusive, and only one of $T$ or $U$ is observed. We observe $T$ if the event appears earlier than censoring $(T\leq U)$, or we do not observe $T$ if censoring happens earlier than the event $(U<T)$. An observation $i$ in time-to-event data consists of $(t,\delta,\mathbf{x})$, where $t$ is the event time or censoring time, depending on which comes first, $\delta$ is the censoring indicator, which is 1 if event time is observed and 0 if the observation is censored, and $\mathbf{x}$ is the vector of covariates. To put it differently, $t=min\{T,U\}$ and $\delta=\mathbf{I}(T\leq U)$.

To perform likelihood-based inference on time-to-event data, one should consider both the case of censored and complete observations, such that the likelihood becomes

\small
\begin{equation}
L=\mathrm{Pr}(T=t,U>t|\mathbf{x})^{\delta}\mathrm{Pr}(T>t,U=t|\mathbf{x})^{1-\delta},
\end{equation}
\normalsize

\noindent which yields a computationally difficult expression, due to the two joint probability distribution functions. A normal assumption made to simplify this expression is the assumption of independent and non-informative censoring (as defined in \cite{EmuraChen2018}): 

\begin{itemize}
\item Independent censoring: Event time and censoring time are independent given the covariates
\item Non-informative censoring: The censoring distribution does not involve any parameters related to the distribution of the survival times
\end{itemize}

\noindent In real world situations, a censoring is usually non-informative if it is independent, and in the rest of the paper we  assume that independent censoring implies non-informative censoring. Note that the independent censoring assumption states that $T$ and $U$ are conditionally independent given $\mathbf{x}$. In other words, even if there exists dependency between $T$ and $U$, when the covariates contain all information about the dependency, the independent censoring assumption holds. This specific case is not explored further in this paper, as most real-world situations rarely provide all the necessary information in the covariates. 

Under the assumption of independent censoring, the likelihood function can be rewritten as

\begin{equation}
\label{eq:separation}
\begin{split}
L&=[\mathrm{Pr}(T=t|\mathbf{x})\mathrm{Pr}(U>t|\mathbf{x})]^{\delta}[\mathrm{Pr}(T>t|\mathbf{x})\mathrm{Pr}(U=t|\mathbf{x})]^{1-\delta}. \\
&= [f_T(t|\mathbf{x})S_U(t|\mathbf{x})]^{\delta}[S_T(t|\mathbf{x})f_U(t|\mathbf{x})]^{1-\delta} \\
&=[f_T(t|\mathbf{x})^{\delta}S_T(t|\mathbf{x})^{1-\delta}][f_U(t|\mathbf{x})^{1-\delta}S_U(t|\mathbf{x})^{\delta}],
\end{split}
\end{equation}

\noindent where $S_T(t|\mathbf{x})=\mathrm{Pr}(T>t|\mathbf{x})$, $f_T(t|\mathbf{x})=-dS_T(t|\mathbf{x})/dt$, $S_U(t|\mathbf{x})=\mathrm{Pr}(U>t|\mathbf{x})$, and $f_U(t|\mathbf{x})=-dS_U(t|\mathbf{x})/dt$. When assuming independent censoring, $f_U(t|\mathbf{x})^{1-\delta}S_U(t|\mathbf{x})^{\delta}$ is unrelated to the event time. Hence, the likelihood function is simplified to

\begin{equation} \label{eq:simple}
\begin{split}
L &\propto f_T(t|\mathbf{x})^{\delta}S_T(t|\mathbf{x})^{1-\delta} \\
&=f_T(t|\mathbf{x})^{\delta}(1-F_T(t|\mathbf{x}))^{1-\delta},
\end{split}
\end{equation}

\noindent which only depends on the probability density function and cumulative distribution function of $T$. Although, if the assumption of independent censoring does not hold, the separation in Eq. \eqref{eq:separation} does not hold. In this case, the likelihood function is expressed as

\begin{equation}
\label{eq:likelihood_dependent}
\begin{split}
L&=\mathrm{Pr}(T=t,U>t|\mathbf{x})^{\delta}\mathrm{Pr}(T>t,U=t|\mathbf{x})^{1-\delta}\\
&=\bigg\{-\frac{\partial}{\partial y}\mathrm{Pr}(T>y,U>t|\mathbf{x})\bigg\rvert_{y=t}\bigg\}^{\delta}\\ 
 & \; \; \; \; \cdot\bigg\{-\frac{\partial}{\partial z}\mathrm{Pr}(T>t,U>z|\mathbf{x})\bigg\rvert_{z=t}\bigg\}^{1-\delta},
\end{split}
\end{equation}

\noindent where the joint survival functions $\mathrm{Pr}(T>y,U>t|\mathbf{x})$ and $\mathrm{Pr}(T>t,U>z|\mathbf{x})$ depend on both the random variables $T$ and $U$. This makes $T$ non-identifiable without making further assumptions on the censoring mechanism \citep{tsiatis1975nonidentifiability}. However, by assuming a specific structure on the dependence between $T$ and $U$, as done with copula theory, $L$ can be written in terms of the marginals, as we will see in Section \ref{sec:clayton_copula}. Before going into details on how the copula function is integrated into the likelihood in Section \ref{sec:copulaXBoost}, we also describe the two other parts which our methodology is based on, the Accelerated Failure Time model in subsection Section \ref{sec:aft} and gradient boosting in Section \ref{sec:xgb}. 

\subsection{Clayton Copula}
\label{sec:clayton_copula}

Copulas are functions used to describe the dependency between random variables, and were introduced by \cite{Sklar1959}. The theory of copulas is based on Sklar's theorem, which states that any multivariate joint distribution can be modelled by its marginal distribution and a copula function,

\begin{equation}
F(x_1,...,x_D)=C(F_1(x_1),...,F_D(x_D)),
\end{equation}

\noindent where $F$ is a joint multivariate cumulative distribution function, $C$ is the copula function and $F_d, d=1,\cdots,D,$ is the marginal cumulative distribution function for the random variable $X_d$.  Sklar's theorem also relates a joint survival function $S(x_1,...,x_D)$ with its marginal survival functions $S_d(x_d)$ in an analogues way as  a joint distribution function $F(x_1,...,x_D)$ with its margins $F_d(x_d)$ \citep{georges2001multivariate,nelsen1995copulas,kaishev2007modelling}. This means that the joint survival function of the event time and censoring time $S_{T,U}(t,u|\mathbf{x})=\mathrm{Pr}(T>t,U>u|\mathbf{x})$  can be expressed in terms of their marginals $S_T(t)=\mathrm{Pr}(T>t|\mathbf{x})$, $S_U(u)=\mathrm{Pr}(U>u|\mathbf{x})$ and a suitable bivariate copula $C$,

\begin{equation}
\label{eq:joint_survival}
S_{T,U}(t,u|\mathbf{x}) = \mathrm{Pr}(T>t,U>u|\mathbf{x})= C(S_T(t|\mathbf{x}),S_U(u|\mathbf{x})).
\end{equation}

\noindent When taking a copula function of the survival functions instead of the cumulative functions, it is referred to as the \textit{survival copula} of $T$ and $U$. 

A copula function $C: [0,1]\times [0,1]\mapsto [0,1]$ must satisfy certain conditions in order to be valid:

\begin{enumerate}
\item $C(a,0)= C(0,b)=0$, $C(a,1)=a$, and $C(1,b) = b$ for $0\leq a\leq 1$ and $0\leq b\leq 1$,
\item $C$ is non-decreasing in every argument,
\end{enumerate}

\noindent which ensures that $C$ is the joint cumulative distribution function of two random variables with uniform marginals. One copula that satisfies these conditions is the Clayton copula, one of the most prominent Archimedean copulas. A bivariate Archimedean copula is defined as

\begin{equation}
C_{\theta}(v_1,v_2)=\phi^{-1}_{\theta}\big(\phi_{\theta}(v_1)+\phi_{\theta}(v_2)\big),
\end{equation}

\noindent where the function $\phi_{\theta}$, the generator of the copula, is continuous and strictly decreasing and $\theta$ is a parameter that describes the dependency strength between the two random variables $v_1$ and $v_2$.  The Clayton copula has

\begin{equation}
\label{eq:clayton}
\phi_{\theta} = \frac{t^{-\theta}-1}{\theta}
\end{equation}

as generator with inverse function

\begin{equation}
\label{eq:inv_clayton}
\phi^{-1}_{\theta}(t)=(t\theta+1)^{-\sfrac{1}{\theta}},
\end{equation}

\noindent which satisfies conditions 1 and 2 as long as $\theta > 0$. An increasing $\theta$ leads to a higher dependency, while if $\theta\rightarrow0$, $v_1$ and $v_2$ goes towards independence. 

The Clayton copula is the simplest copula among the Archimedean copulas: It does not require any logarithmic or exponential operation and has only one parameter governing the dependency strength. The Clayton copula has already been exhaustively explored within the area of survival analysis, as it has an historical important role in the introduction of copulas within this field \citep{clayton1978model,oakes1982model,hsu1996assessing, shih1998goodness}. One reason for this is the mathematical simplicity of the Clayton copula, which makes it preferable when modelling the dependency between event and censoring time compared to other copulas. The Clayton copula is  also especially interesting within time-to-event prediction as its survival version has an asymmetric structure and exhibits greater dependence in the higher tail. An increasing dependence between $T$ and $U$ with higher values seems reasonable in many real-world applications. An example is the dataset FRICTION described in Section \ref{sec:real_data}, where censoring will happen almost exclusively at higher friction values. Also for medical use cases, such as the GBSG2 dataset described in \ref{sec:real_data}, it is likely to assume that censoring mechanisms happening later in the disease progress (e.g. death or drop-out) are more related to the disease progress, while censoring mechanism happening early are more generally caused by randomness.

If a dataset is known to have a dependency structure with a greater dependency in the lower tail, it would be beneficial to explore the use of a Gumbel copula. If the dependency structure is known to be symmetric, a Frank copula would be suitable. However, due to the dependency structure of the Clayton Copula and its mathematical simplicity and  important role within survival analysis, the Clayton Copula is explored in this work, limited to the case when $\theta > 0$ to satisfy conditions 1 and 2. 

\subsection{Accelerated Failure Time Model} \label{sec:aft}

The most popular regression models for handling censored data is arguably the Cox Proportional Hazards (Cox PH) model, highly utilized within biomedicine, followed by the Accelerated Failure Time (AFT) model, which is more used within reliability theory. The former assumes a multiplicative effect of the covariates on the hazard rate. The main reason why many prefer the Cox PH model is because it is semi-parametric, and estimation and inference is possible without making any assumptions of the baseline distribution. However, Cox PH is based on the proportional hazard assumption, which means that the ratio of hazards for two individuals is constant over time. If this assumption does not hold, the Cox PH might give biased predictions or it might not converge.  

The AFT model, on the other hand, assumes that the effect of the covariates is to accelerate or decelerate the event time by some constant. In this way, the parameters of the AFT model are easier to interpret than for the Cox PH, since they directly measure the effect of the covariates on the event time \citep{wei1992accelerated}. Another advantage with the AFT model is that it does not need the proportional hazard assumption. The main disadvantage with the AFT model is that it is a parametric model, and one have to make an assumption on the baseline event function. For a more exhaustive discussion on the differences between the two models, we refer the reader to \cite{wei1992accelerated} and \cite{orbe2002comparing}.

In this work, we choose to work with the AFT model, as this gives an easier access to the predicted  event time instead of modelling hazard functions. The original Accelerated Failure Time model assumes a linear model for the logarithm of the event time,

\begin{equation}\label{eq:aft}
\log (T)=\mathbf{\beta} \mathbf{X}+\mathcal{E},
\end{equation}

\noindent where $\mathbf{\beta}$ is the coefficients, $\mathbf{X}$ the covariate matrix and $\mathcal{E}=\sigma_Z Z$ is the
error, that follows a baseline distribution Z, and has mean 0 and variance $\sigma_Z^2$. Typical choices for probability distributions for $Z$ is Gumbel, normal and logistic, which in terms leads to baseline functions for $T$ to be Weibull, log-normal and log-logistic, respectively. Since we do not want to limit our model to assume linear effects of the covariates, we generalize the AFT model by substituting the linear function of the covariates by a general function

\begin{equation} \label{eq:gen:aft}
\log(T)=h(\mathbf{X})+\sigma_Z Z,
\end{equation}

\noindent where $h(\mathbf{X})$ captures the effect of $X$ on the response, and can be estimated from any model in broad generality.

\subsection{Gradient boosting and XGBoost}\label{sec:xgb}

As mentioned in the introduction, from a machine learning point of view this paper focuses on the gradient boosting approach. As in any supervised learning setting, the goal here is to construct the function $h(\mathbf{x})$ that, given the vector of covariates $\mathbf{x}$, provides a good estimate $\hat{y} = h(\mathbf{x})$ of the response $y$, which in our case is $\log(T)$. The goodness of the estimate is evaluated in terms of a loss function, $\mathcal{L}(y, \hat{y})$. When working with time-to-event data, $y$ is formed by the event time $t$ and the censoring indicator $\delta$. The idea behind gradient boosting is rather simple: Starting from the null model, the algorithm iteratively improves by fitting a base learner, we will use a statistical tree, to the negative gradient of the loss function computed at the current model. Basically, at each iteration the algorithm seeks the fastest way to minimize the loss function starting from the current point and fits the base learner to capture it. The ensemble of all the fits is the final model. Peculiar of gradient boosting, these single improvements are made artificially small through a penalization parameter, to control the speed of the minimization process and therefore better explore the model space. See Algorithm \ref{alg:gboost} for a schematic view of boosting.

\begin{algorithm}
\caption{{Gradient boosting} }
\begin{algorithmic}
\item 1. Initialize $h(\mathbf{x})$;
\item 2. Update the model by, for $k = 1, \dots, K$,
\begin{itemize}
\item[2.1] compute the negative gradient of the loss function;
\item[2.2] obtain the improvement $h_k(\mathbf{x})$ by fitting the base learner on the negative gradient;
\end{itemize}
\item 3. Aggregate the results, $h(\mathbf{x}) = \sum_{k = 0}^K h_k(\mathbf{x})$.
\label{alg:gboost}
\end{algorithmic}
\end{algorithm}

To implement the boosting approach we use the eXtreme Gradient Boosting algorithm \citep{ChenGuestrin2016}, with statistical trees as the base learner. A notable characteristic of XGBoost is that it uses, in addition to numerous computational tricks, a second order approximation of the loss function in its computations, to speed up the procedure. In particular, it turns the updating step 2 of Algorithm \ref{alg:gboost} into the minimization problem
  
\begin{equation}\label{eq:obj}
\begin{split}
h_k(\boldsymbol x) = \text{argmin}_{h_k(\boldsymbol x)} \left[ \mathcal{L}\left(y, \sum_{j=0}^{k-1}h(\boldsymbol x)^{[j]}\right) \right.  \\
\left. + g_1 h_k(\boldsymbol x) + \tfrac{1}{2} g_2 h^2_k(\boldsymbol x) \vphantom{\int_1^2} \right] + \Omega(h_k(\boldsymbol x)),
\end{split}
\end{equation}\\

where $\sum_{j=0}^{k-1}h(\boldsymbol x)^{[j]}$ is the current (i.e., up to the previous step) estimate of the model,
\begin{equation*}
\begin{split}
g_1 = \left. \frac{\partial}{\partial h(\boldsymbol x)} \mathcal{L}(y,h(\boldsymbol x))\right|_{h(\boldsymbol x) = \sum_{j=0}^{k-1}h(\boldsymbol x)^{[j]}}, \\ g_2 = \left. \frac{\partial^2}{\partial h(\boldsymbol x)^2} \mathcal{L}(y,h(\boldsymbol x))\right|_{h(\boldsymbol x) = \sum_{j=0}^{k-1}h(\boldsymbol x)^{[j]}},
\end{split}
\end{equation*}
\noindent and $\Omega(h_k(\boldsymbol x))$ is a penalty term that penalizes the base learner complexity, in this case the tree. $h_k(\boldsymbol x)$ is indeed the statistical tree that solves Eq. \ref{eq:obj}. Noticeably, its fitting, i.e. the computation of the split points and the leaf weights, only depends on the loss function through $g_1$ and $g_2$, and its complexity penalised by 
\begin{equation}
\Omega(h_k(\boldsymbol x)) = \gamma T_k + \frac{1}{2}\lambda ||w_k||^2,
\end{equation}
\noindent where $\gamma$ controls the penalty for the number of tree leafs $T$ and $\lambda$ the magnitude of the weights $w$, with $||\cdot||$ denoting the L$_2$ norm. As we mentioned above,  it is important for the boosting algorithm that the update on each step does not improve the model too much.

Note that the algorithm is very general, and merely requires the specification of the right loss function to apply XGBoost to a specific problem (and provide the algorithms with the related first and second derivatives). The square loss $\mathcal{L}(y, \hat{y}) = (y - \hat{y})^2$ is usually implemented for Gaussian regression, while the negative log-likelihood of the binomial distribution for classification problems. Therefore, our main task here is to derive a loss function that works for data involving dependent censoring. This is covered in the next section.

\section{Clayton-boost}\label{sec:copulaXBoost}

Derivation of the appropriate loss function starts with modelling the joint survival function in Eq. \eqref{eq:likelihood_dependent} in terms of the marginal survival functions. Due to Sklar's theorem (Eq. \ref{eq:joint_survival}), we know that the copula function yields a valid model for the joint survival functions, such that the likelihood can be expressed as

\small
\begin{equation}
\label{eq:likely}
\begin{split}
L = \bigg\{-\frac{\partial}{\partial y}C_{\theta}(S_T(y|\mathbf{x}),S_U(t|\mathbf{x}))\bigg\rvert_{y=t}\bigg\}^{\delta}\\
\cdot\bigg\{-\frac{\partial}{\partial z}C_{\theta}(S_T(t|\mathbf{x}),S_U(z|\mathbf{x})\bigg\rvert_{z=t}\bigg\}^{1-\delta}.
\end{split}
\end{equation}
\normalsize

Taking the log of the likelihood, $\ell = \log(L)$, and expressing Eq. \eqref{eq:likely} in terms of a Clayton copula (Eq. \ref{eq:clayton}) yields

\small
\begin{equation}
\label{eq:copula_loss}
\begin{split}
&\ell = \delta\log\bigg\{-\frac{\partial}{\partial y}\bigg(\big(S_T(y|\mathbf{x})^{-\theta}+S_U(t|\mathbf{x})^{-\theta}-1\big)^{-\sfrac{1}{\theta}}\bigg)\bigg\rvert_{y=t}\bigg\}\\
&+(1-\delta)\log\bigg\{-\frac{\partial}{\partial z}\bigg(\big(S_T(t|\mathbf{x})^{-\theta}+S_U(z|\mathbf{x})^{-\theta}-1\big)^{-\sfrac{1}{\theta}}\bigg)\bigg\rvert_{z=t}\bigg\},
\end{split}
\end{equation}
\normalsize

\noindent which is on an computationally easier format, as the log-likelihood depends only on the marginalized survival functions and the dependency parameter $\theta$.  The expression can be further simplified by the use of the chain rule and basic rules for logarithms, and by using the fact that $S(t|\mathbf{x})=1-F(t|\mathbf{x})$ and $\frac{\partial}{\partial y}F(y|\mathbf{x})=f(y|\mathbf{x})$, to become

\small
\begin{equation}
\begin{split}
\ell =&-\big(1+\tfrac{1}{\theta}\big)\log\big((1-F_T(t|\mathbf{x}))^{-\theta}+(1-F_U(t|\mathbf{x}))^{-\theta}-1\big)\\
&-\big(1+\theta)\delta\log\big(1-F_T(t|\mathbf{x}_i)\big)
+\delta\log\big(f_T(t|\mathbf{x})\big)\\
&-\big(1+\theta\big)\big(1-\delta\big)\log\big(1-F_U(t|\mathbf{x})\big)+\big(1-\delta\big)\log\big(f_U(t|\mathbf{x})\big).
\end{split}
\end{equation}
\normalsize

Maximizing this log-likelihood function is the same as minimizing its negative, so the loss function to be minimized is

\small
\begin{equation}
\label{eq:jointloss}
loss =\big(1+\tfrac{1}{\theta}\big)\log\big((1-F_T(t|\mathbf{x}))^{-\theta}+(1-F_U(t|\mathbf{x}))^{-\theta}-1\big)+g(\delta,t).
\end{equation}
\normalsize

where

\small
\begin{equation*}
\label{eq:dividedloss}
  g(\delta,t) =
    \begin{cases}
      & \big(1+\theta)\log\big(1-F_T(t|\mathbf{x})\big)-\log\big(f_T(t|\mathbf{x})\big),\; \; \delta=1\\
      & \big(1+\theta\big)\log\big(1-F_U(t|\mathbf{x})\big)-\log\big(f_U(t|\mathbf{x})\big),\; \; \delta=0.
    \end{cases}       
\end{equation*}
\normalsize

\noindent The first part of the loss function is independent of $\delta$, while the contribution of $g(\delta,t)$ depends on $\delta$ and is either affected by the conditional distribution of $T$ if we observe the true event time, or the conditional distribution of $U$ if the time is censored.

Unfortunately, the distribution functions of $T$ and $U$, conditional on the covariates, are rarely known. This can be addressed by assuming a model for the effect of the covariates on the event and censoring time, and assuming a known  baseline distribution, as done with an  Accelerated Failure Time model in Eq. \eqref{eq:gen:aft}. To write the loss function in Eq. \eqref{eq:jointloss} in terms of the known baseline distribution $Z$, we apply the transformation $\omega(Z)=\exp(h(\mathbf{x})+\sigma_Z Z)$ according to the AFT model, and the fact that $F_T\big(t\big)=F_Z\big(\omega^{-1}(t)\big)$ and $f_T(t) = f_Z\big(\omega^{-1}(t)\big)\frac{d}{dt}\omega^{-1}(t)$ since $t=\omega(Z)$ and $\omega$ is an increasing, continuous, monotone function on $\mathcal{Z}=\{z:f_Z(z)>0\}$. Applying this transformation yields:

\small
\begin{equation}
\begin{split}
F_T\big(t\big)=F_Z\bigg(\frac{\log(t)-h(\mathbf{x})}{\sigma_Z}\bigg)=F_Z\big(s(t)\big)\\
f_T\big(t\big)=f_Z\bigg(\frac{\log (t)-h(\mathbf{x})}{\sigma_Z}\bigg)\frac{1}{\sigma_Z t}=\frac{f_Z\big(s(t)\big)}{\sigma_Z t},
\end{split}
\end{equation}
\normalsize

\noindent where $s(t)=\frac{\log (t)-h(\mathbf{x})}{\sigma_Z}$. This is a similar approach to the independent censoring AFT implementation in XGBoost \citep{barnwal2020survival}, which we later refer to as \textit{Std-boost}. However, since our model also include information about the censoring distribution, we apply the same transformation for the distributions of the censoring times:

\small
\begin{equation}
\begin{split}
F_U\big(t\big)=F_V\bigg(\frac{\log (t)-h(\mathbf{x})}{\sigma_V}\bigg)=F_V\big(r(t)\big)\\
f_U\big(t\big)=f_V\bigg(\frac{\log (t)-h(\mathbf{x})}{\sigma_V}\bigg)\frac{1}{\sigma_V t}=\frac{f_V\big(r(t)\big)}{\sigma_V t},
\end{split}
\end{equation}
\normalsize

\noindent where $F_V$ and $f_V$ are the CDF and PDF of the random variable $V$, which is  distributed with mean 0, $\sigma_V^2$ is the variance of the error term and $r(t)=\frac{\log (t)-h(\mathbf{x})}{\sigma_V}$.

Substituting these transformations into Eq. \eqref{eq:jointloss} gives the final loss function

\small
\begin{equation}
\label{eq:final_loss}
\begin{split}
loss =&\big(1+\tfrac{1}{\theta}\big)\log\big((1-F_Z(s(t)|\mathbf{x}))^{-\theta}\\
&+(1-F_V(r(t)|\mathbf{x}))^{-\theta}-1\big)+g(\delta,t)
\end{split}
\end{equation}
\normalsize

where

\small
\begin{equation*}
\label{eq:final_g}
  g(\delta,t) =
    \begin{cases}
      & \big(1+\theta)\log\big(1-F_Z(s(t)|\mathbf{x})\big)\\
      &-\log\big(f_Z(s(t)|\mathbf{x})\frac{1}{\sigma_Z t}\big),\qquad \quad\delta=1\\
      & \\
      & \big(1+\theta\big)\log\big(1-F_V(r(t)|\mathbf{x})\big)\\
      &-\log\big(f_V(r(t)|\mathbf{x})\frac{1}{\sigma_V t}\big),\qquad \quad \delta=0.
    \end{cases}       
\end{equation*}
\normalsize

The loss function can be incorporated into many statistical and machine learning frameworks, as long as they allow a customized loss function. In this work, we illustrate the effect of the loss function by incorporating it into a gradient boosting approach described in Section \ref{sec:xgb}, which gives us Clayton-boost, the boosting algorithm for dependent censoring. 

As seen in Eq. \eqref{eq:final_loss}, Clayton-boost requires the user to provide the baseline functions for both the event distribution $Z$ and the censoring distribution $V$, as well as their standard deviations $\sigma_Z$ and $\sigma_V$. When used in practice, these distributions can be empirically inferred from the data by looking at the sampling distributions for both the uncensored and censored  event times. Another approach is to test different distributions and standard deviations on the dataset by using cross validation, as done in \cite{barnwal2020survival}, and choosing the distribution which provides the best performance. The latter approach can be done when no preliminary information is possible to retrieve from the data, or just to minimize the manual data exploration. However, in our experiments on real data in section \ref{sec:real_data}, we choose to instead assume a baseline distribution empirically from the sampling distributions of the training data, which is shown in detail in Appendix \ref{sec:appendix_histograms} . In the implementation of Clayton-boost, the available distributions for the baseline functions are Gumbel, normal and logistic.

Another parameter that has to be provided in the loss function is the dependency parameter $\theta$. If no prior knowledge about this parameter exists, it can be tuned as a hyper parameter when training the statistical model, e.g., with cross validation. This approach was used in the application on real datasets in Section \ref{sec:real_data}. However, in our simulation study in Section \ref{sec:simulation}, the true dependency $\theta$ is known, and we choose to use the true $\theta$ also for the dependency parameter in the Clayton-boost algorithm. 

\section{Applications}\label{applications}

\subsection{Evaluation Criteria}\label{sec:eval}

To evaluate our novel algorithm and to contrast it to existing approaches, we measure both its calibration and discrimination ability. By looking at the difference between the predicted and the actual outcomes, and at the ability of correctly ranking the predictions in terms of their event times, respectively, these two criteria fully capture the prediction ability of a model \citep{SteyerbergAl2010, DebinAl2014}. We measure them by computing the concordance index \citep{HarrellAl1982}, the mean absolute error and visualizing the calibration plot.

\subsubsection{Concordance index} In general, the concordance index (c-index) measures the number of concordant pairs of observations among all possible pairs. When applied to time-to-event data analysis, the term ``concordant'' means that the observations with  a shorter predicted event time (or higher risk) in fact have a shorter observed event time. The measurement is complicated by the presence of censoring, which makes some pairs ``unusable''. Modifications of the basic index have been developed to tackle this issue, two notable being those of \cite{UnoAl2011} and \cite{GerdsAl2013}. However, both versions belong to the class of inverse of the probability of censoring weighted (IPCW) estimators, and assume independent censoring given the covariates. In this paper, we use the version of \cite{harrell1996multivariable} from the Lifelines Package \citep{pilon}, since it does not assume independent censoring directly through the IPCW modification. Nevertheless, Harrell's c-index is not a perfect evaluation criteria  either, as it is dependent on the censoring distribution, which might introduce some bias when $\theta \neq 0$ \citep{brentnall2018use}. In addition,  Harrell's c-index is known to be overoptimistic (biased upwards) with increasing amount of censoring \citep{UnoAl2011, han2017comparing,kim2015estimation}.

\subsubsection{Mean Absolute Error} To account for the problems which arises with the use of c-index at the presence of dependent censoring, we also evaluate  the models by the use of mean absolute error (MAE). For the simulation studies, we provide the MAE directly, since the true event time is known also for the censored observations. However, for the real-world datasets, the true event time is unknown for censored observations, and we provide instead the event MAE calculated only on the uncensored observations. While this measure does not provide a general estimation of the model error, as it ignores all the censored observations in the test data, it it still useful in estimating the performance of the model for the observed event times. Since event MAE provides additional information about how much predictions deviate from the truth, and not only if they were ranked correctly, it is a criterion sometimes used for evaluating time-to-event predictions together with concordance index (e.g. \cite{matsuo2019survival,kiaee2015relevance}).  In addition, in some applications the performance of the model may also only be relevant for the uncensored observations.  For example, in our third real world example FRICTION, the prediction of the time-to-event (the available friction coefficient) is only useful when the event occurs, and the model does not need to be evaluated in situation in which normal braking is sufficient.

\subsubsection{Calibration plot} This graphical tool allows evaluating whether the predicted event times are close to the actual ones. The observations are divided in bins, and the predicted and observed proportions of events within the bins are measured. These two measures serve as coordinates for adding points to a graph. The closer the line connecting the points is to the line with intercept 0 and slope 1, the better the calibration. See \cite{VancalsterAl2019} for more details on this kind of plot and on the importance of assessing the calibration of a prediction model.

\paragraph{} Note that classical measures for censored data, such as the IPCW Brier score \citep{GrafAl1999, gerds2006consistent}, are not suitable as evaluation criteria in our setting, as they are designed for independent censoring. The IPCW Brier score is heavily dependent on the estimates of the censoring distribution \citep{KvammeBorgan2019}, which are commonly estimated assuming independent censoring using e.g. the Kaplan-Meier estimator. When the censoring times are related to the covariates or the event times, the IPCW Brier Score is no longer valid. Due to the limitations of all evaluation criteria in the presence of dependent censoring, the combination of all three should be used to evaluate the performance of the models. 

\subsection{Simulations} \label{sec:simulation}
\subsubsection{Data generating mechanism}

Simulating data for the evaluation of our proposed model is not straightforward, as we have to simulate both the copula function between $T$ and $U$ with a given $\theta$, as well as the marginal distributions for both $T$ and $U$, and simulate the effect of covariates on $T$ and $U$. We approach this problem by first assuming a stochastic model for the event time

\begin{equation} \label{eq:T}
T = f(X_1,...,X_m)\cdot R_1
\end{equation}

\noindent where $f$ is a deterministic function of the covariates, $m$ is the number of covariates and $R_1$ is a stochastic variable with a given distribution. Then, we set the censoring distribution as a product of a constant $c$ and a stochastic variable $R_2$

\begin{equation} \label{eq:U}
U = cR_2
\end{equation}

If we assume that $R_1\sim Weibull(\lambda,k)$, then the conditional distribution of $T$ given $f(X_1=x_1,...,X_m=x_m)$ also follows a Weibull distribution: $T\sim Weibull(f(x_1,...,x_m)\cdot \lambda,k)$, where $f(x_1,...,x_m)$ affects the scaling factor of the distribution. If we assume the same Weibull distribution for $R_2$ as for $S$, the percentage of censoring is easily controlled by varying the constant $c$; $U\sim Weibull(c\lambda,k)$ to give the censoring distribution a higher or lower scale than the event distribution. A low value for $c$ will provide a higher censoring percentage.

The structure of $T$ (Eq. \eqref{eq:T}) makes it possible to choose a variety  of effects of the covariates on the event time, as long as $f(X_1,...,X_n) > 0$ when $S$ is Weibull distributed. To ensure a positive scaling, we set $f$ as an exponential function of the covariates,

\begin{equation} \label{eq:f}
f(X_1,...,X_m) = e^{h(X_1,...,X_m)},
\end{equation}

\noindent making $T$ the AFT model given in Eq. \eqref{eq:gen:aft}

\begin{equation}
\log (T)=h(X_1,...,X_n)+\mathcal{E}.
\end{equation}

\noindent where $\mathcal{E} = log(R_1)$, which makes $\mathcal{E}$ a Gumbel Minimum variable with $\mu=-log(\lambda)$, $\beta = \frac{1}{k}$ and standard deviation $\sigma_Z=\beta\frac{\pi}{\sqrt{6}}$. To simulate a scenario containing non-linear functions, we create a function $h$ consisting of eight uniform covariates with interaction effects, and add two noise covariates non-related to the event time: 

\begin{equation} \label{eq:h}
h(X_1,...,X_{10})=X_1X_2+\frac{1}{2}X_3^3+X_4X_5+\frac{4}{5}e^{-X_6}+X_7\mathrm{sin}(2X_8)
\end{equation}

\noindent where $X_1,...,X_{10}\sim Uniform(0,1)$. The coefficients of the covariates are selected to ensure that all relevant variables have a somewhat equal contribution to $h$. 

After choosing the marginal distributions for $T$ and $U$ by setting $R_1$, $R_2$, $h(X_1,...,X_m)$ and $c$, the dependency between $T$ and $U$ is created by inducing rank correlation according to uniform distributed Clayton copula variables. This is done in the following way: Draw two uniform distributed random variables with a dependency according to a Clayton copula and dependency parameter $\theta$. Uniform Clayton copula variables can be simulated using the connection between the inverse of the Clayton's generator function \eqref{eq:inv_clayton} and the Laplace transformation of a Gamma-distribution \citep{hofert2008sampling}. The Marshall, Olkin algortihm  \citep{marshall1988families} then becomes:

\begin{algorithm}
\caption{Simulating data from the Clayton Copula}\label{alg:cap}
\begin{algorithmic}
\item Input $\theta > 0$
\item Draw $K \sim Gamma(\frac{1}{\theta},1)$
\item For $i \in [1,2]$
{\setlength\itemindent{20pt} \item Draw $X_i \sim Uniform(0,1)$}
{\setlength\itemindent{20pt} \item Set $W_i \gets (1-\frac{log X_i}{K})^{-\frac{1}{\theta}}$}
\item Return $W_1, W_2$
\end{algorithmic}
\end{algorithm}
 
\noindent The dependency between the two uniform variables $W_1$ and $W_2$ can be induced into the variables $T$ and $U$ based on rank correlation \citep{iman1982distribution}. Induced rank correlation is easy to use,  distribution free and  preserves the  marginal distributions of the original variables $T$ and $U$. Using this method, we want to recreate the same rank correlation of $T$ and $U$ as $W_1$ and $W_2$ by permuting the values of $T$ and $U$. The result is a new set $T_{W}$ and $U_{W}$ where the marginal distributions are the same as $T$ and $U$, while the rank correlation is exactly the same as between $W_1$ and $W_2$. 

The use of induced rank correlation was quite popular after its introduction in 1982. In our simulations, we use the same rank correlation inducing regime as the original paper \citep{iman1982distribution} in two dimensions. However, we alter the approach by drawing the uniform variables from a Clayton copula instead of a Gaussian copula. This novel approach has, to the best of the authors' knowledge, not been pursued before. One benefit of using an Archimedean copula here is that the true dependency $\theta$ will be retained between the two new variables $T_{W}$ and $U_{W}$, since the Clayton dependency parameter $\theta$ is directly related to rank correlation by the Kendall's Tau $\tau$ \citep{EmuraChen2018}:
 
 \begin{equation}
\tau = \frac{\theta}{\theta+2}.
\end{equation}

\noindent Since the rank correlation between $T_{W}$ and $U_{W}$ exactly matches  the rank correlation between $W_1$ and $W_2$, the $\theta$ between $T_{W}$ and $U_{W}$ will also exactly  match the $\theta$ between $W_1$ and $W_2$. After this permutation, $T_{W}$ and $U_{W}$ are used as the true event time and censoring time respectively, and the observed time is the minimum value, $t=\mathrm{min}\{T_{W},U_{W}\}$. In addition, the covariates $X_m, m \in [0,10]$ has to be permuted according to the same rank as $T_{W}$ and $U_{W}$.

One thing to note about the simulation settings is that the simulated model does not directly match the assumed dependency in the Clayton-boost loss function (Eq. \ref{eq:final_loss}). Clayton-boost assumes a Clayton dependence between the survival functions of $T$ and $U$ conditional on the covariates, as seen in Eq. \eqref{eq:joint_survival}. In our simulation setup, the induced dependence between the survival functions of $T$ and $U$ are done unconditionally of the covariates. If the data would be simulated to have a conditional dependence, the effect of the covariates would alter the dependence structure such that it does not correspond to a Clayton copula, which would also not match the dependency assumed in Clayton-boost. Therefore, the former settings were chosen. However, the authors found it natural that the simulated data does not correspond directly with the loss function in Clayton-boost, as this would often be the case in real-world settings. To put it differently, Clayton-boost will get a potential disadvantage in the simulations, since the simulated model does not directly match the Clayton-boost model. 

\subsubsection{Settings}

\begin{table*}
\centering
\caption{The relationship between Kendall's $\tau$ and the values for $\theta$ used in simulation study 1}
{\tabcolsep8pt
\begin{tabular}{@{}llllllllll@{}}\toprule 

$\theta$ &  $1e^{-10}$ & 1 & 2 & 3 & 4 & 5 & 6 & 7 & 8\\ \toprule 

Kendall's  $\tau$ & $5e^{-11}$ & 0.33 & 0.50 & 0.60 & 0.67 & 0.71 & 0.75 & 0.78 & 0.80\\ \toprule 
\label{tab:kendall}
\end{tabular}}
\end{table*} 

\begin{table*}
\centering
\caption{The Archimedean copulas used for comparison of models' performance in simulation study 3. $D_1$ is the Debye function of the first kind}
{\tabcolsep8pt
\begin{tabular}{@{}llll@{}}\toprule

Copula & $C_{\theta}(u,v)$ & $\theta \in$ & Kendall's $\tau$ \\ \toprule

Clayton & $(u^{-\theta}+v^{-\theta}-1)^{-\tfrac{1}{\theta}}$ & $ [-1,\infty){\setminus}\{0\}$ & $\frac{\theta}{\theta+2}$  \\
Gumbel & $\mathrm{exp}[-((-\log (u))^{\theta}+(-\log (u))^{\theta})^{\frac{1}{\theta}}]$ & $[1,\infty)$ & $\frac{\theta-1}{\theta}$  \\
Frank & $-\frac{1}{\theta}\log [1+\frac{(e^{-\theta u}-1)(e^{-\theta v})-1)}{e^{-\theta}-1}]$ & $\mathbb{R}{\setminus}\{0\}$ & $1+\frac{4}{\theta}(D_1(\theta)-1)$  \\ 
Independent & $uv$ &  & 0  \\\toprule 
\label{tab:copulas}
\end{tabular}}
\end{table*} 

\cite{EmuraChen2018} studied the estimation bias of Cox regression in the presence of dependent censoring, which showed that the bias increased  with both increasing $\theta$ and percentage censoring. We will evaluate how Clayton-boost performs under difficult censoring dependency by simulating data under these to situations in simulation study 1 and 2. Another point of interest is how well the Clayton-boost model works on other types of dependency structures than Clayton induced dependencies, which is explored in Simulation study 3. 

\textbf{Simulation study 1:} To study the effect of increasing $\theta$, the model is evaluated on simulated data containing 1000 training samples and 1000 test samples, where both $R_1,R_2\sim Weibull(1,3)$ and the event times, censoring time and covariates are drawn according to Eq. \eqref{eq:T}, \eqref{eq:U}, \eqref{eq:f} and \eqref{eq:h}. We set $c=1.49$, which gives approximately 50\% censoring, and vary $\theta$ between $1e^{-10}$ (approx. 0) and $8$. The corresponding Kendall's $\tau$ to the $\theta$'s are given in Table \ref{tab:kendall}. The simulation is repeated 20 times, and the average performance is reported.

\textbf{Simulation study 2:}  To study the effect of percentage censoring, the model is evaluated with the same simulation set up as Simulation study 1, but by varying $c$ between 0.89 (approx. 90\% censoring) and 2.06 (approx 10\% censoring), while keeping $\theta$ fixed at 3. The simulation is repeated 20 times, and the average performance is reported.

\textbf{Simulation study 3:}  To explore the effect of different dependency structures, the model is evaluated at data simulated from four different Archimedean copulas given in Table \ref{tab:copulas}. While the Clayton copula has a stronger dependency in the lower tail, the Gumbel copula has a stronger dependency in the higher tail, and the Frank copula has a symmetric dependency. The independent copula has no dependency. Example scatter plots of $T$ and $U$ with the induced copulas on their survival functions are shown in Figure \ref{fig:scatter} in Appendix. The model performance is evaluated by inducing the different copulas dependencies between $T$ and $U$ with $c=1.2$. This corresponds to approximately 70\% censoring, as according to the results from simulation 2 given in Section \ref{sec:results} and Figure \ref{fig:percentage}, the effect of censoring should be quite clear at this percentage. In addition, we set the values for $\theta$ such that all the models will have the same Kendall's $\tau$ of 0.6 between $T$ and $U$ (excluding the independent copula, which will always have a Kendall's $\tau$ of 0). The simulation is repeated 20 times, and the average performance is reported.

\subsubsection{Competitors}
The performance of Clayton-boost is evaluated by comparing it to the two standard approaches for time-to-event predictions, namely AFT and Cox PH. \textit{Std-AFT} is an implementation of the classical AFT model (eq. \ref{eq:aft}) in Python, and is part of the package Lifelines \citep{pilon}. This model assumes independent censoring, by taking advantage of the likelihood function given in eq. \eqref{eq:simple}. As mentioned in Section \ref{sec:aft}, the AFT model assumes that the covariates either accelerate  or decelerate the event time by some constant, and is a parametric model where a baseline function is assumed. The Lifelines package contains three different functions for fitting either a Weibull, log-normal and log-logistic distribution. 

\textit{Cox} is an implementation of Cox  PH model from the Lifelines package, which also rely on the assumption of independent censoring through the likelihood function Eq. \eqref{eq:simple}. As mentioned in Section \ref{sec:aft}, the Cox model assumes that the covariates either accelerate or decelerate the hazard rate, and is based in the proportional hazard assumption.

In addition to the two common time-to-event prediction methods, we compare our model to an AFT-boosting model.
\textit{Std-boost} is a prior implementation of AFT in XGBoost \citep{barnwal2020survival}, which also assumes independent censoring through Eq. \eqref{eq:simple} and is based on the AFT model (Eq. \ref{eq:aft}). However, this method uses a boosting procedure to optimize the likelihood.  

All compared models (including Clayton-boost) were trained and tuned using 2-fold-cross-validation, with one tunable parameter for fair comparison. The details regarding the set up can be found in Appendix \ref{apx:simulation_setup}, and the code for the simulation studies can be found here: \url{github.com/alimid/clayton_boost/tree/main/simulation_studies}.

\subsubsection{Results} \label{sec:results}

\begin{figure*}
\centering
\begin{minipage}{.5\linewidth}
\centering
\subfloat[MAE]{\label{res:a}\includegraphics[scale=.35]{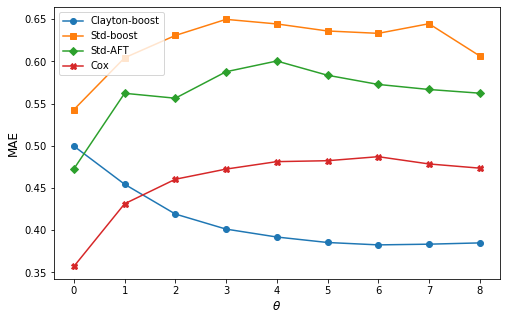}}
\end{minipage}%
\begin{minipage}{.5\linewidth}
\centering
\subfloat[C-Index]{\label{res:b}\includegraphics[scale=.35]{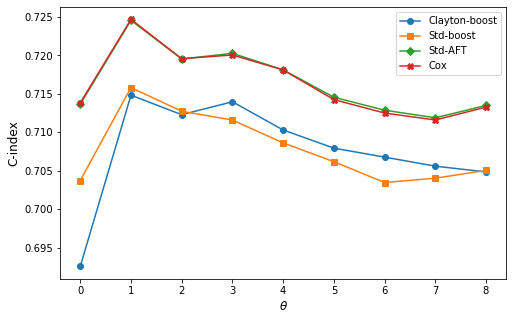}}
\end{minipage}\par\medskip

\caption{Model performance as a function of $\theta$ with approx. 50\% censoring}
\label{fig:results_theta}
\end{figure*}

\begin{figure*}
\centering
\begin{minipage}{.33\linewidth}
\centering
\subfloat[$\theta=1e^{-10}$]{\label{cal_theta:a}\includegraphics[scale=.35]{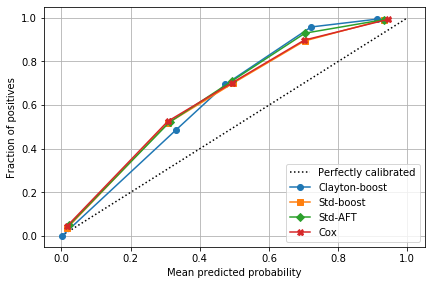}}
\end{minipage}%
\begin{minipage}{.33\linewidth}
\centering
\subfloat[$\theta=4$]{\label{cal_theta:b}\includegraphics[scale=.35]{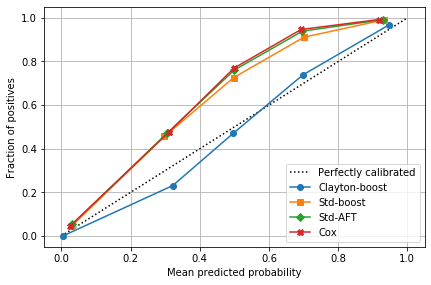}}
\end{minipage}
\begin{minipage}{.33\linewidth}
\centering
\subfloat[$\theta=8$]{\label{cal_theta:c}\includegraphics[scale=.35]{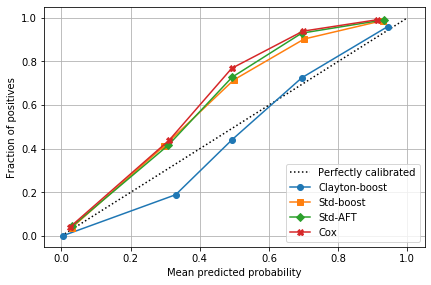}}
\end{minipage}\par\medskip

\caption{Calibration plots for the models under data simulated from different $\theta$ with approx. 50\% censoring. The full figure with all calibration plots is given in Figure \ref{fig:app_theta} in the appendix}
\label{fig:cal_theta}
\end{figure*}

\textbf{Simulation study 1}. The performance of Clayton-boost and the three competing models are evaluated at different values of $\theta$ and the results are shown in terms of MAE and c-index in Figure \ref{fig:results_theta} and in terms of calibration plots in Figure \ref{fig:cal_theta}.

Figure \ref{res:a} shows a clear trend that an increasing $\theta$ provides higher MAE for all the comparing models, with the trend flattening out for higher dependencies. This is natural, as the the derivative of the effect of $\theta$ on correlation is decreasing, as shown in Table \ref{tab:kendall}. Clayton-boost, on the other hand,  actually lowers the  MAE as the dependency increases. The reason for this is probably that the information available about the true event time given by the censoring time increases with increasing $\theta$, and Clayton-boost is capable of using this information. Clayton-boost has a higher MAE than Cox and std-aft at independence ($\theta = 1e^{-10} \approx 0$), but after dependency is added, it performs better than both AFT-based models, and as $\theta$ increases to 2, it has the lowest MAE of all the models. 

The c-index of the models in Figure \ref{res:b} also shows a decreasing performance when the dependency increases, apart from going from $\theta=1e^{-10}$ to $\theta=1$, where a bias is introduced when adding dependency ($\theta \neq 0$) due to the c-index being dependent on the censoring distribution, as discussed in Section \ref{sec:eval}.  However, we can assume that an equal bias is added to all the models, such that they can be compared to each other. We then observe that, in the case of ranking the event times, Clayton-boost follows the same trend as the other models, and that both boosting models at all times have a lower c-index than both std-AFT and Cox.

The calibration plots in Figure \ref{fig:cal_theta} show that the competing models in general underestimate the mean event probability. In other words, the models overestimate the event times. Estimation bias in the presence of higher percentage censoring is a well-known problem when using maximum likelihood estimation methods \citep{esteve1990relative, pan1998nonparametric, shen2020nonparametric,moeschberger1985comparison, barrajon2020effect}. In the specific case of the AFT models, the problem originates in the first derivative of the likelihood function, which always has a positive value for right censored observations. This means that the AFT model tends to provide too high predictions of the event times if the percentage of right censoring is considerable. Therefore, a bias in the predictions for all models at 50\% censoring, also when the censoring is independent, is not a surprise. However, the plots show a small trend that the overestimation increases for higher event times with increasing $\theta$, while the overestimation lowers for lower event times. This corresponds to the structure of the Clayton copula, which affects higher tail values.  

In Figure \ref{fig:cal_theta}, Clayton-boost also underestimates the mean event probability for independence, but as $\theta$ increases, it decreases the prediction bias and almost overlaps with the diagonal line. However, Clayton-boost underestimate the event times slightly for lower event times when $\theta$ increases highly. Nevertheless, with dependency present, Clayton-boost seems to be much better calibrated than the competing models. The full figure with all calibration plots is given in Appendix \ref{apx:calibration_plots}.

The results from simulation study 1 shows that the models are indeed affected by dependency strength. However, Clayton-boost manages much better than the competing models to correct overestimation and MAE in the presence of Clayton-induced dependence, while both std-AFT and Cox remain better at ranking the predictions. 

\begin{figure*}
\centering 
\label{fig:percentage}
\begin{minipage}{.5\linewidth}
\centering
\subfloat[MAE]{\label{per:a}\includegraphics[scale=.35]{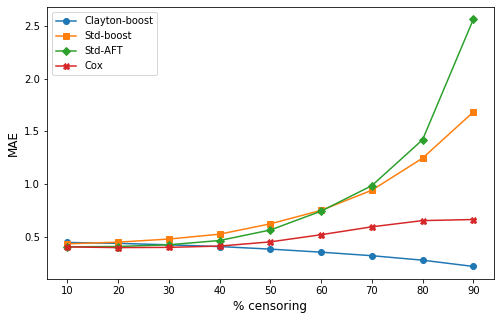}}
\end{minipage}%
\begin{minipage}{.5\linewidth}
\centering
\subfloat[C-index]{\label{per:b}\includegraphics[scale=.35]{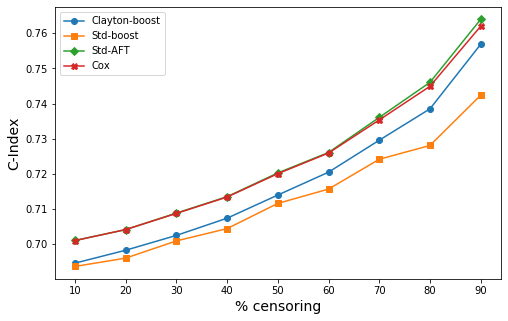}}
\end{minipage}\par\medskip

\caption{Model performance as a function of percentage censoring with $\theta=3$}
\label{fig:percentage}
\end{figure*}

\begin{figure*}
\begin{minipage}{.33\linewidth}
\centering
\subfloat[$10\%$]{\label{cal_percentage:a}\includegraphics[scale=.35]{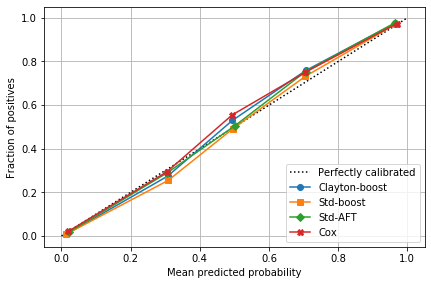}}
\end{minipage}%
\begin{minipage}{.33\linewidth}
\centering
\subfloat[$50\%$]{\label{cal_percentage:b}\includegraphics[scale=.35]{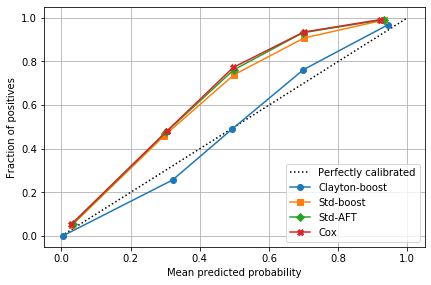}}
\end{minipage}
\begin{minipage}{.33\linewidth}
\centering
\subfloat[$90\%$]{\label{cal_percentage:c}\includegraphics[scale=.35]{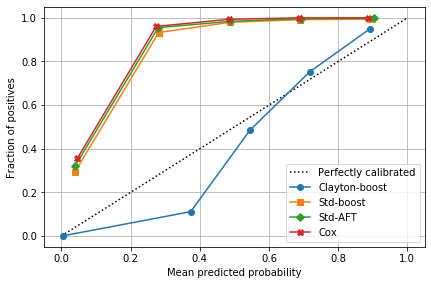}}
\end{minipage}\par\medskip

\caption{Calibration plots for the models under data simulated from different percentage censoring with $\theta=3$. The full figure with all calibration plots is given in Figure \ref{apx:cal_theta} in the appendix}
\label{fig:cal_percentage}
\end{figure*}

\textbf{Simulation study 2}. The performance of the Clayton-boost and the three competing models are evaluated at different percentage censoring and the results are shown in terms of MAE and c-index in Figure \ref{fig:percentage} and in terms of calibration plots in Figure \ref{fig:cal_percentage}.

Figure \ref{fig:percentage} shows a very clear trend, that the MAE for all the competing models increase as the percentage of censoring increases. Clayton-boost, on the other hand, has a quite horizontal line with a small decreasing trend. Clayton-boost starts with the highest  MAE of the models at 10\% censoring, but outperforms std-boost as the censoring goes beyond 20\%, and has the lowest MAE at 40\%. At 90\% censoring, the difference between the models' performances are large. Regarding the c-index, we observe that it increases with the percentage of censoring for all the models. However, this is not necessarily an effect of the models' performance, but rather the known phenomena that Harrell's concordance index is over-optimistic if the amount on censoring is high \citep{UnoAl2011, han2017comparing,kim2015estimation}.
Therefore, it is difficult to infer any effect from amount of censoring on the models' performance through the c-index. 

\begin{figure*}[!btp]
\centering
\begin{minipage}{.45\linewidth}
\centering
\subfloat[Clayton $\theta=3$]{\label{box:a}\includegraphics[scale=.25]{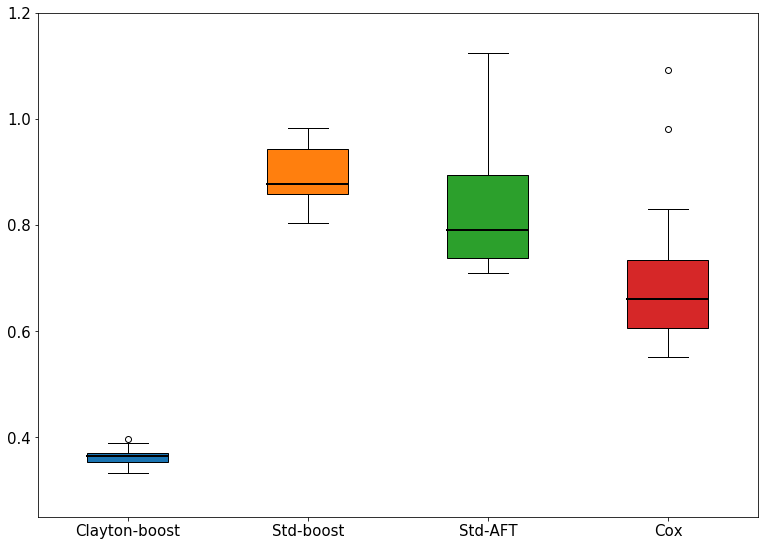}}
\end{minipage}%
\begin{minipage}{.45\linewidth}
\centering
\subfloat[Gumbel $\theta=2.5$]{\label{box:b}\includegraphics[scale=.25]{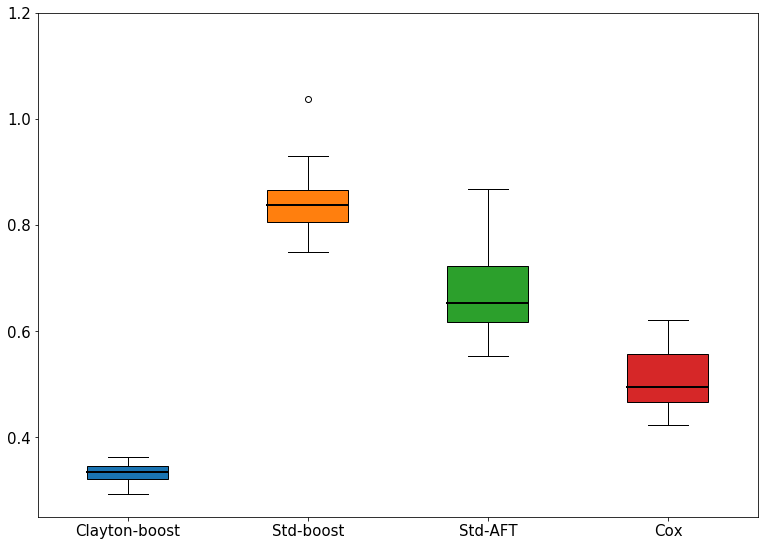}}
\end{minipage}\par\medskip

\begin{minipage}{.45\linewidth}
\centering
\subfloat[Frank $\theta=7.5$]{\label{box:c}\includegraphics[scale=.25]{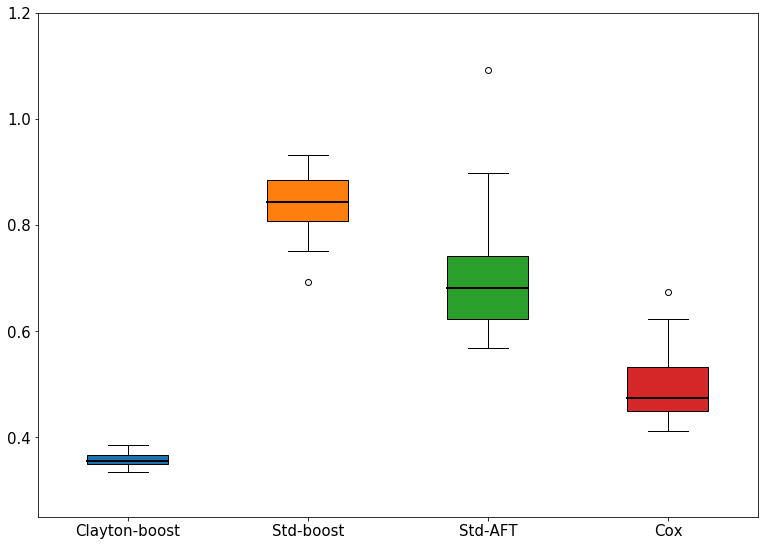}}
\end{minipage}%
\begin{minipage}{.45\linewidth}
\centering
\subfloat[Independent]{\label{box:d}\includegraphics[scale=.25]{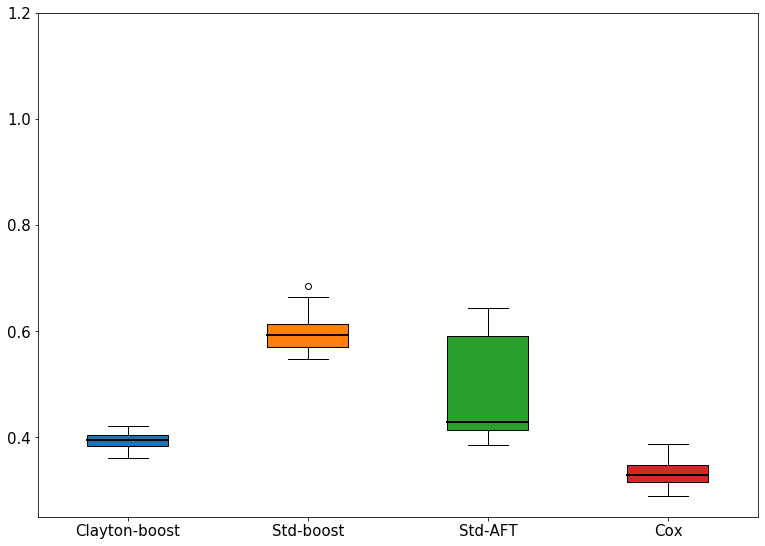}}
\end{minipage}\par\medskip

\caption{Box plots over models' MAE for the four induced copulas with approx. 70\% censoring}
\label{fig:box}
\end{figure*}
 
The calibration plots in Figure \ref{fig:cal_percentage} show that at only 10\% censoring, all models seem very well calibrated. However, as the censoring increases to 50\%, the competing models overestimate the event times, in a similar manner as simulation study 1. When the censoring becomes as high as 90\%, the competing models' overestimation becomes very high. In fact, they seem to predict that only 30\% of the events happens when close to 100\% actually happened. This result corresponds which previous research, stating that the effect of falsely assuming independent censoring is substantial at higher percentage censoring and very low for low percentage censoring \citep{Lagakos1979} Clayton-boost, on the other hand, stays well calibrated also for 50\% censoring. However, we see that for very high percentage censoring, it lies close to the diagonal line for higher event times, but seems to underestimate the even times for lower values. However, the general calibration is again much better for Clayton-boost than the models assuming independent censoring.

\begin{figure*}
\centering
\begin{minipage}{.45\linewidth}
\centering
\subfloat[Clayton $\theta=3$]{\label{box_c:a}\includegraphics[scale=.25]{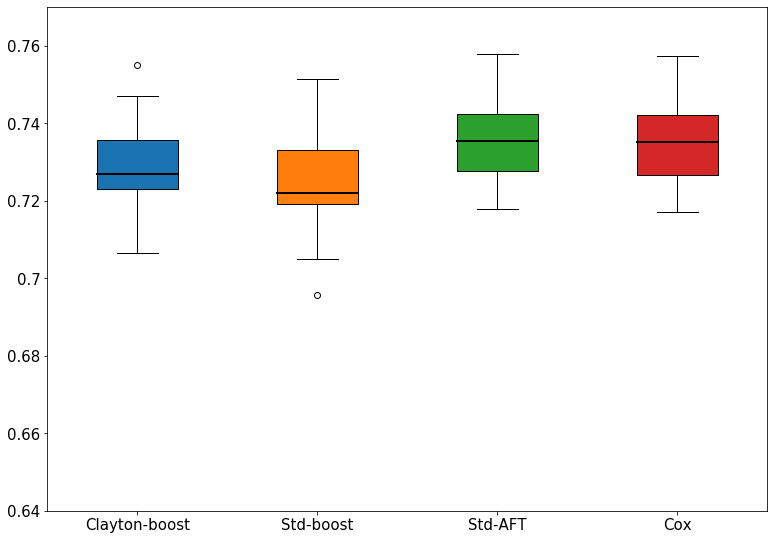}}
\end{minipage}%
\begin{minipage}{.45\linewidth}
\centering
\subfloat[Gumbel $\theta=2.5$]{\label{box_c:b}\includegraphics[scale=.25]{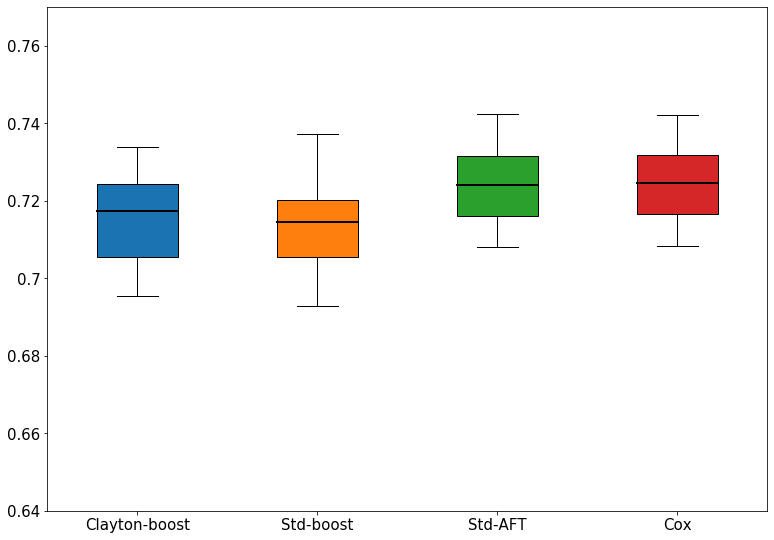}}
\end{minipage}\par\medskip

\begin{minipage}{.45\linewidth}
\centering
\subfloat[Frank $\theta=7.5$]{\label{box_c:c}\includegraphics[scale=.25]{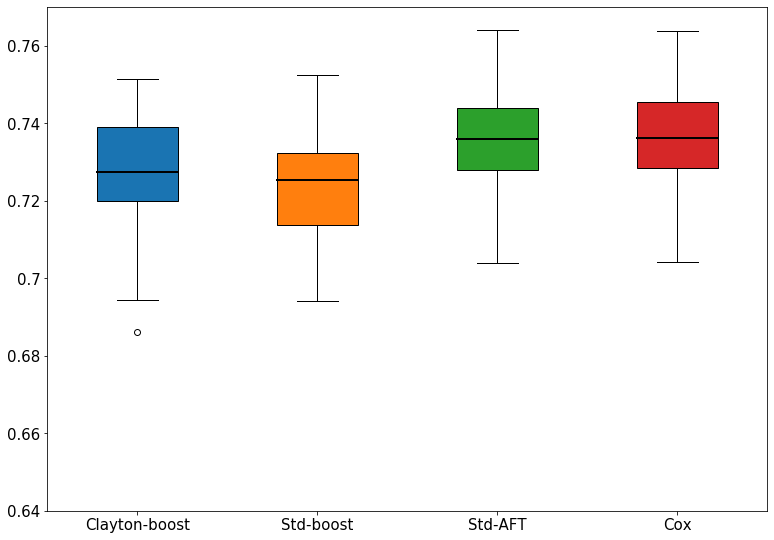}}
\end{minipage}%
\begin{minipage}{.45\linewidth}
\centering
\subfloat[Independent]{\label{box_c:d}\includegraphics[scale=.25]{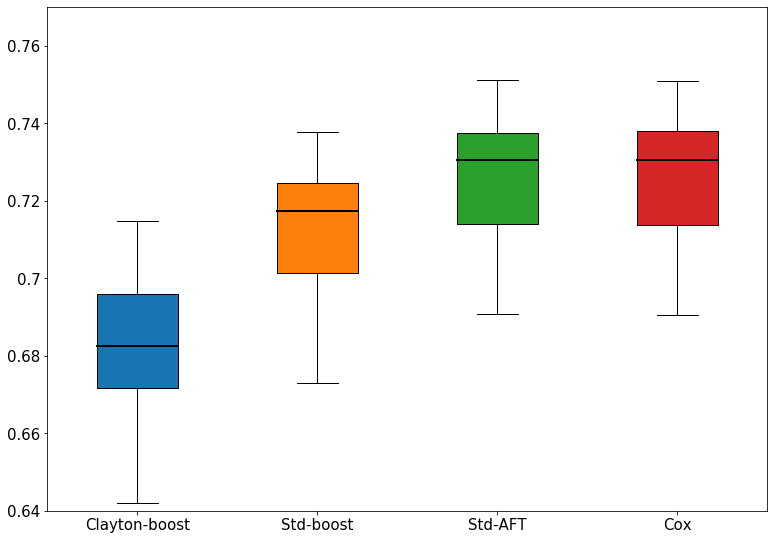}}
\end{minipage}\par\medskip

\caption{Box plots over models' c-index for the four induced copulas with approx.  70\% censoring}
\label{fig:box_c}
\end{figure*}

\begin{figure*}
\centering
\begin{minipage}{.5\linewidth}
\centering
\subfloat[Clayton $\theta=3$]{\label{cal_diff:a}\includegraphics[scale=.35]{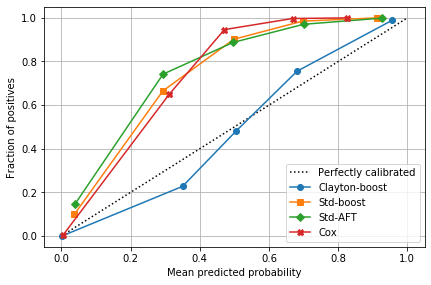}}
\end{minipage}%
\begin{minipage}{.5\linewidth}
\centering
\subfloat[Gumbel $\theta=2.5$]{\label{cal_diff:b}\includegraphics[scale=.35]{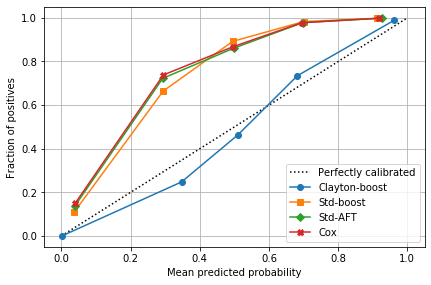}}
\end{minipage}\par\medskip

\begin{minipage}{.5\linewidth}
\centering
\subfloat[Frank $\theta=7.5$]{\label{cal_diff:c}\includegraphics[scale=.35]{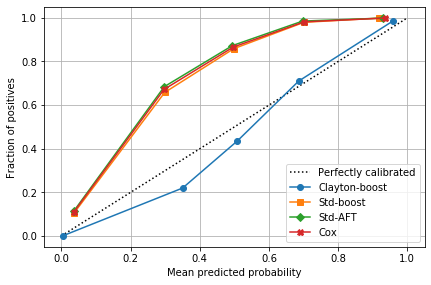}}
\end{minipage}%
\begin{minipage}{.5\linewidth}
\centering
\subfloat[Independent]{\label{cal_diff:d}\includegraphics[scale=.35]{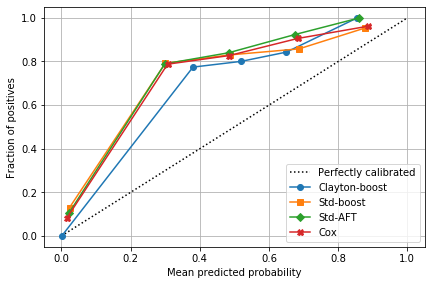}}
\end{minipage}\par\medskip

\caption{Calibration plots for the four induced copulas with approx. 70\% censoring}
\label{fig:cal_diff}
\end{figure*}

\textbf{Simulation study 3}. The results of 20 simulations for the four induced copulas are shown as box plots in Figure \ref{fig:box} and \ref{fig:box_c} and as calibration plots in Figure \ref{fig:cal_diff}. We observe that Clayton-boost has a stable, low MAE for all the dependent censoring structures, not only for the Clayton induced one, and strongly outperforms the competing models. For the independent copula, however, the competing models also perform reasonably well, and the Cox model performs better than Clayton-boost. When it comes to c-index, the performance varies between the different models. Std-boost has the lowest performance on all of the copulas both in terms of MAE and c-index, with the exception of the independent copula, where Clayton-boost has the lowest c-index. Cox has a marginally better c-index than the other models for all copulas.

The calibration plots shows that Clayton-boost is much better calibrated for all the copulas with dependence than the competing models, which overestimate the event times. However, under the independent copula, Clayton-boost behaves similarly to the competing models. Note that the results for the independent copula differs from Simulation Study 1 ( $\theta=1e^{-10}$) as we here have a higher percentage censoring (70\% vs 50\%).

The results show that the Clayton-boost model can be used for other dependence structures than Clayton, when the goal is to predict the  time to event, as it outperforms the other models in regards to MAE. However, when there really is no dependency between $T$ and $U$, other models could be preferred. Additionally, if the problem of interest is to only rank the observations correctly in time, and not estimate the true event time, Clayton-boost does not seem to provide any advantages. An incorrect assumption of independent censoring  seems to mostly affect the time estimate, not the ranking of times.

\subsection{Real data} \label{sec:real_data}

\begin{table*}[!b]
\centering
\caption{Description of the datasets used in the experiment. $n$ is the number of observations and $m$ is the number of covariates}
{\tabcolsep8pt
\begin{tabular}{@{}lllll@{}}\toprule 

Dataset & n & m & Censoring & Task \\ \toprule 

GSBG2 & 686 & 9 & 56\% & Days to recurrence-free survival \\

DD & 1808 & 33 & 19\% & Years of governing body \\

FRICTION & 15154 & 106 & 95\% & Friction coefficient \\ \toprule 
\label{tab:data}
\end{tabular}}
\end{table*} 

In addition to the simulated data, the performance of Clayton-boost is evaluated on three different real-world datasets:

\begin{itemize}
\item GBSG2 is a dataset from the German Breast Cancer Study Group, containing observations from a study of 686 women used for comparing time to recurrence-free and overall survival between different treatment modalities \citep{schumacher}. The censoring mechanism is loss of life, which is quite related to overall survival: If a patient passes away, the person concerned was probably not close to gaining recurrence-free survival. The proportions of observed and non-observed survival is quite even, with 56\% censoring.

\item DD was originally a study to classify political regimes as Democracy or Dictatorship, but is here used to predict the duration of a country's governing body. The censoring mechanism is quite different than in the GBSG2-dataset, as it is a combination of administrative censoring and random censoring. The administrative censoring comes from the end-of-study in 2008, where all regimes were censored at this point, since we do not know how much longer the regimes last. When entrance time is part of the covariates, administrative censoring is independent given the covariates. In addition to the administrative censoring there is another censoring mechanism present in the dataset. Whether or not this random censoring is independent of governing time is difficult to infer from the data. However, both censoring mechanisms are rarer, as only 19\% of the dataset is censored. Both the GSBG2 and the DD datasets were loaded from the Python package Lifelines \citep{pilon}. 

\begin{figure*}
\centering
\begin{minipage}{.5\linewidth}
\centering
\subfloat[GSBG2]{\label{cal:a}\includegraphics[scale=.35]{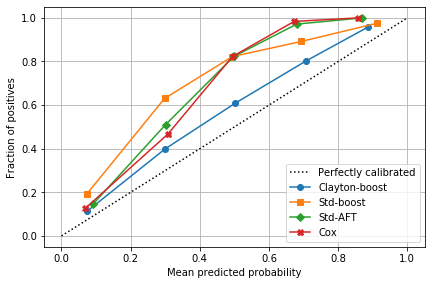}}
\end{minipage}%
\begin{minipage}{.5\linewidth}
\centering
\subfloat[DD]{\label{cal:b}\includegraphics[scale=.35]{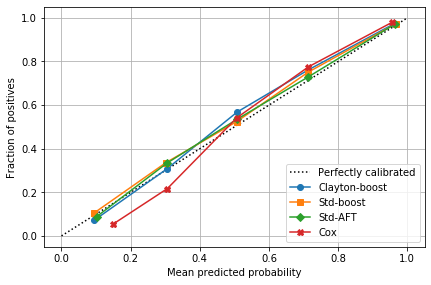}}
\end{minipage}\par\medskip
\centering
\subfloat[FRICTION]{\label{cal:c}\includegraphics[scale=.35]{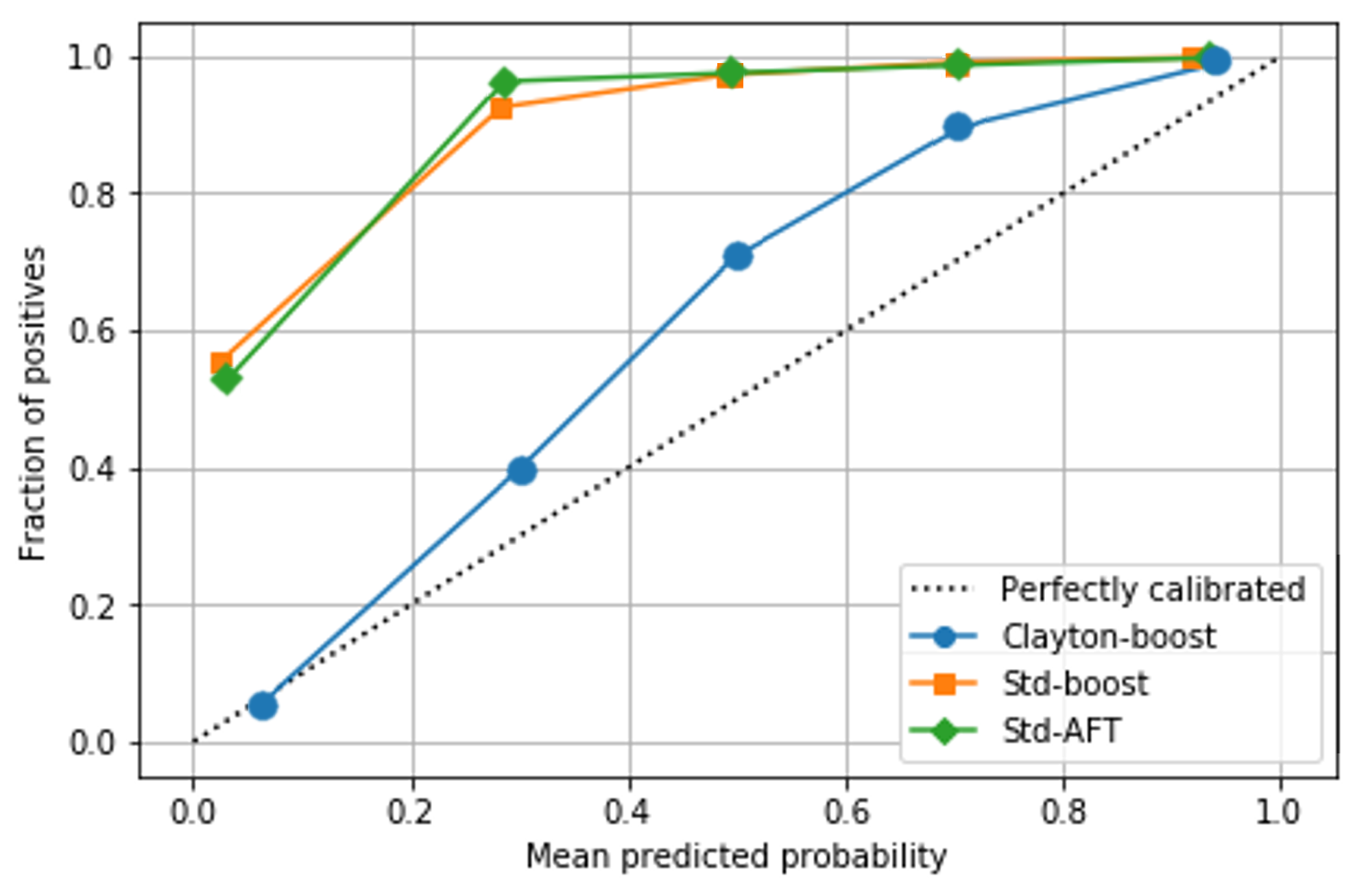}}

\caption{Calibration plots for the real-world datasets}
\label{fig:cal}
\end{figure*}

\begin{table*}[!b]
\centering
\caption{Performance of the models on the real-world datasets.}
{\tabcolsep8pt
\begin{tabular}{@{}lllllllll@{}}\toprule 

 & \multicolumn{2}{c}{Clayton-boost} & \multicolumn{2}{c}{Std-boost} & \multicolumn{2}{c}{Std-AFT} & \multicolumn{2}{c}{Cox} \\ 

Dataset & C-index & MAE & C-index & MAE & C-index & MAE & C-index & MAE \\ \toprule

GBSG2 & \textbf{0.705} & \textbf{746} & 0.660 & 1208 & 0.678 & 1323 & 0.679 & 875 \\
DD & 0.684 & \textbf{2.970} & \textbf{0.695} & 2.972 & 0.652 & 3.996 & 0.651 & 3.719 \\
FRICTION & 0.799 & \textbf{0.047} & \textbf{0.838} & 0.356 & 0.835 & 1.148 & - & - \\ \toprule 
\label{tab:reg_results}
\end{tabular}}
\end{table*} 

\item In addition to the public datasets, we showcase another use of time-to-event emerging from the world of engineering and risk analysis.  In the dataset FRICTION, the response variable is not time, but a continuous variable which is right censored, namely the tire-to-road friction coefficient. The dataset contains weather data and runway reports, which are used to predict the runway surface friction (tire-to-road friction coefficient), to support in safer airplane landings. The censoring mechanism is that a landing is not friction limited (does not use all the available friction on the runway), where one only knows the amount of used friction, which becomes a lower bound of the available friction. As most airplane landings happens during safe runway conditions, the data is highly censored at 95\%. As a lot of the same variables can be used to describe both if a landing is friction limited and the actual friction coefficient (as shown in \cite{midtfjord2021}), there are good reasons for assuming dependent censoring. Further information about this dataset can be found in \cite{midtfjord2021} and \cite{midtfjord2020}.  Note  that these two papers use airplane landings from ten years, while this work only considers data for one winter season (2017/2018), to shorten computing time.  A summary of the three datasets is given in Table \ref{tab:data}.
\end{itemize}

By looking at the calibration plot for the GBSG2 data in Figure \ref{fig:cal}, we see that all of the other models in general overestimate the true survival time, i.e. consequently underestimate the survival probability.  Clayton-boost, on the other hand, manages to remove a lot of this consequent bias, and lies closer to the diagonal line. The performance evaluation metrics in Table \ref{tab:reg_results}, shows that both c-index and event MAE is best for the Clayton-boost model, followed by the Cox model. However, the event MAE is really high for both of the independent AFT based models (above 1200 days), showing that these highly miss on the true time to overall survival. If the problem of interest is to predict the time until survival, these models are not very useful. Hence, it is clear that the GBSG2 data benefited from having the censoring dependency taken into account.

The calibration plot for the DD data in Figure \ref{cal:b} is quite different from the calibration plots for the two other datasets, as all the models lies quite close to the diagonal line. The models do not experience the same large overestimation of the event times as with the GBSG2 data. And as seen in Table \ref{tab:data}, both boosting models outperform the classical, statistical models, while the performance of Clayton-boost and std-boost are very similar. In other words, we don't observe any large benefits from modelling censoring dependency for the DD dataset. This could be an indication that the independent censoring assumptions holds for this data, both for the administrative and random censoring. However, the main reason is probably the low percentage censoring of 19\%. As seen in our simulations in Figure (\ref{per:a}), with this low censoring percentage less than 20\%, even a strong dependency of $\theta = 3$ does not affect the models performances much. The similar behavior on the real-world dataset DD amplifies the result that the prediction bias is highly dependent on percentage censoring, and for lower percentage censoring, assuming independent censoring does not seem to affect model performance particularly. 

In the FRICTION dataset, it was seen in \cite{midtfjord2021} that a lot of the same covariates were related to both the friction coefficient and the censoring mechanism, which would violate the independent censoring assumption, if we assume that not all information about the dependency is given in the covariates. As seen if Figure \ref{fig:cal}, the methods making the independent censoring assumption heavily overestimate the friction coefficient, and the overestimation is largest for the lower frictions. This is an unfortunate effect, as the lower frictions represent the dangerous conditions for flight landings and are the ones important to identify. The Clayton-boost algorithm also overestimates the friction coefficients, but in a much lower manner than the competing models. And it is reverted: The bias is smaller for the lower frictions, and higher for increasing frictions. The latter would not be as important for this use case, as this would mean not dangerous situations, and modelling the precise friction coefficient is not important. Looking at the results in Table \ref{tab:reg_results}, Clayton-boost has a lower c-index than the competing  models, but a much lower MAE (0.047 compared to 0.356 and 1.148). It is easily seen that modelling the dependency had a large impact on this dataset. One reason for this could be the high percentage of censoring (95\%). As seen in the simulations (Figure \ref{fig:percentage}), this high amount of censoring greatly impacts the performance of the models. Note: The Cox model did not work for this particular dataset, due to convergence problems. There might be several reasons for this as explained in the Lifelines documentation\footnote{\url{https://lifelines.readthedocs.io/en/latest/Examples.html}}, e.g. the high occurrence of multicollinearity in the dataset.

\section{Conclusion}\label{conclusions}

This paper introduces Clayton-boost, a novel boosting model for performing time-to-event prediction for data involving dependent censoring. The proposed model consists of a loss function which builds upon the accelerated failure time model, uses copulas to model the dependency between the event and censoring distribution and is implemented in a gradient boosting procedure. 

Clayton-boost shows a strong ability to remove prediction bias in the presence of dependent censoring, which the models assuming independent censoring suffer from. This especially applies to data with higher percentage censoring where classical maximum likelihood estimation methods tend to greatly overestimate the event times. Model comparisons on simulated data show that models assuming independent censoring tend to increase prediction bias as dependency strength or percentage censoring increases.  Clayton-boost, on the other hand, seems unaffected by percentage censoring and actually tends to get better as dependency strength increases, probably due to the ability of getting more information from the censored observations. Evaluations at real-world datasets also shows that Clayton-boost outperformed the independent censoring models at higher percentage censoring. 

Being able to use advanced statistical methods, such as gradient boosting, on data with dependent censoring opens many opportunities, as this can often be the case in real-world data. This work showed that there are indeed reasons to be critical about the independent censoring assumption usually made by time-to-event prediction methods, as modelling the dependency could provide considerably higher performance. 

One issue brought to the spotlight in this paper is the need for a proper performance evaluation metric, which is valid also for dependent censoring. The current evaluation metrics provide either biased predictions at the presence of dependent censoring (c-index and Brier Score) or ignores the censored observations (MAE). 

The results in this paper also highlight a main issue with the popular time-to-event prediction methods, namely the increasing prediction bias with increasing percentage censoring, also during independent censoring. We observe that the main advantage of Clayton-boost is its ability to remove much of this bias when the censoring mechanism is dependent on the event times. This makes Clayton-boost a desirable model for real-world time-to-event analysis, as it provides a good performance where the commonly used methods fail.

The Clayton-boost algorithm can be found on this GitHub page: \url{github.com/alimid/clayton_boost}. It is also planned to make a python package for the model within short time. 

\bibliographystyle{cas-model2-names}
\bibliography{Reference2}

\onecolumn
  
\appendix
\section{Appendix}

\subsection{The gradient statistics of the loss function}

\begin{equation}
\frac{\partial loss}{\partial \hat{t}} = (1+\theta)\frac{\big(1-F_Z(s)\big)^{-(1+\theta)}f_Z(s)\frac{\partial s}{\partial \hat{t}}+\big(1-F_V(r)\big)^{-(1+\theta)}f_V(r)\frac{\partial r}{\partial \hat{t}}}{\big(1-F_Z(s)\big)^{-\theta}+\big(1-F_V(r)\big)^{-\theta}-1}+\frac{\partial}{\partial \hat{t}}g(\delta)
\end{equation}

where

\begin{equation*}
  \frac{\partial}{\partial \hat{t}}g(\delta) =
 -(1+\theta)\bigg(\frac{f_W(q)\frac{\partial}{\partial \hat{t}}q}{\big(1-F_W(q)\big)}\bigg)
 -\frac{f_W'(q)\frac{\partial}{\partial \hat{t}}q}{f_W(q)}
 -\frac{\frac{\partial}{\partial \hat{t}}\frac{\partial}{\partial t}q}{\frac{\partial}{\partial t}q},    
\end{equation*}

and 

\begin{gather*}
  W =
    \begin{cases}
      & Z,\; \; \delta=1\\
      & V,\; \; \delta=0,
    \end{cases} \; \; \; \; \; \; q =  \begin{cases}
      & s,\; \; \delta=1\\
      & r,\; \; \delta=0.
    \end{cases}      
\end{gather*} 	

\begin{equation}
\begin{split}
\frac{\partial^2 loss}{\partial \hat{t}^2} =(1+\theta)\frac{\scriptsize
\bigg(1+\theta\bigg)\bigg(1-F_Z(s)\bigg)^{-(2+\theta)}f_Z^2(s)\bigg(\dfrac{\partial}{\partial \hat{t}}s\bigg)^2+\bigg(1-F_Z(s)\bigg)^{-(1+\theta)}\bigg(f_Z'(s)\bigg(\dfrac{\partial}{\partial \hat{t}}s\bigg)^2+f_Z(s)\dfrac{\partial^2}{\partial \hat{t}^2}s\bigg)}{\big(1-F_Z(s)\big)^{-\theta}+\big(1-F_V(r)\big)^{-\theta}-1} \\
+(1+\theta)\frac{\scriptsize\bigg(1+\theta\bigg)\bigg(1-F_V(r)\bigg)^{-(2+\theta)}f_V^2(r)\bigg(\dfrac{\partial}{\partial \hat{t}}r\bigg)^2+\bigg(1-F_V(r)\bigg)^{-(1+\theta)}\bigg(f_V'(r)\bigg(\dfrac{\partial}{\partial \hat{t}}r\bigg)^2+f_V(r)\dfrac{\partial^2}{\partial \hat{t}^2}r\bigg)}{\big(1-F_Z(s)\big)^{-\theta}+\big(1-F_V(r)\big)^{-\theta}-1}\\
-\theta\bigg(\frac{\scriptsize \bigg(1-F_Z(s)\bigg)^{-(1+\theta)}f_Z(s)\dfrac{\partial}{\partial \hat{t}}s+\bigg(1-F_V(r)\bigg)^{-(1+\theta)}f_V(r)\dfrac{\partial}{\partial \hat{t}}r }{\big(1-F_Z(s)\big)^{-\theta}+\big(1-F_V(r)\big)^{-\theta}-1}\bigg)^2+\frac{\partial^2}{\partial \hat{t}^2}g(\delta)
\end{split}
\end{equation}

where

\begin{equation*}
\begin{split}
  \frac{\partial^2}{\partial \hat{t}^2}g(\delta) = &
 -\bigg(1+\theta\bigg)\frac{f_W'(q)\bigg(\dfrac{\partial}{\partial \hat{t}}q\bigg)^2+f_W(q)\dfrac{\partial^2}{\partial \hat{t}^2}q}{1-F_W(q)}
 -\bigg(1+\theta\bigg)\bigg(\dfrac{f_W(q)\dfrac{\partial}{\partial \hat{t}}q}{1-F_W(q)}\bigg)^2\\
 & -\dfrac{f_W''(q)\bigg(\dfrac{\partial}{\partial \hat{t}}q\bigg)^2+f_W'(q)\dfrac{\partial^2}{\partial \hat{t}^2}q}{f_W(q)} 
 +\bigg(\dfrac{f_W'(q)\dfrac{\partial}{\partial \hat{t}}q}{f_W(q)}\bigg)^2
 -\dfrac{\dfrac{\partial^2}{\partial \hat{t}^2}\dfrac{\partial}{\partial t}q}{\dfrac{\partial}{\partial t}q}
 +\bigg(\dfrac{\dfrac{\partial}{\partial \hat{t}}\dfrac{\partial}{\partial t}q}{\dfrac{\partial}{\partial t}q}\bigg)^2.    
 \end{split}
\end{equation*}

\subsection{Simulation set-up}\label{apx:simulation_setup}

\begin{figure}

\begin{minipage}{.5\linewidth}
\centering
\subfloat[Clayton $\theta=3$]{\label{box:a}\includegraphics[scale=.5]{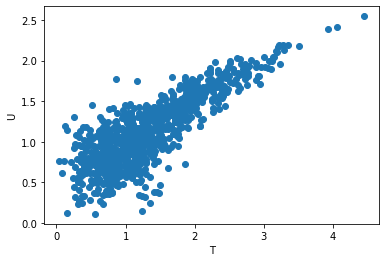}}
\end{minipage}%
\begin{minipage}{.5\linewidth}
\centering
\subfloat[Gumbel $\theta=2.5$]{\label{box:b}\includegraphics[scale=.5]{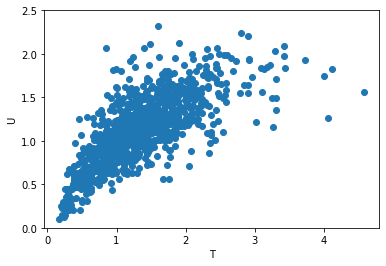}}
\end{minipage}\par\medskip
\begin{minipage}{.5\linewidth}
\centering
\subfloat[Frank $\theta=7.5$]{\label{box:c}\includegraphics[scale=.5]{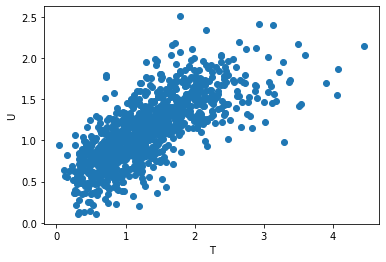}}
\end{minipage}%
\begin{minipage}{.5\linewidth}
\centering
\subfloat[Independent]{\label{box:d}\includegraphics[scale=.5]{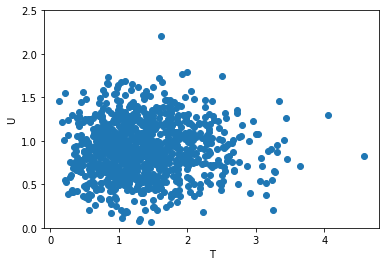}}
\end{minipage}%

\caption{Scatterplot between $T$ and $U$ with different induced copulas and approx. 70\% censoring.}
\label{fig:scatter}
\end{figure}

The full experiments on the simulated data can be found in the Jupyter Notebooks on this GitHub page: \url{https://github.com/alimid/clayton_boost/tree/main/simulation_studies}.

For all the simulated data, the true known baseline distribution for $T$ and $U$, with standard deviations, are known. This also applies for $\theta$. Therefore, this information is given to the Clayton-boost, std-boost and std-AFT algorithms (COX does not use a baseline distribution). 

To create a fair comparison, all the four models were tuned with a cross validation grid search with one tuning parameter. The grid search is optimized on maximizing the c-index, and the settings is shown in Table \ref{tab:cv_sim}.

\begin{table}
\caption{Parameter tuning for the different algorithms.}
{\tabcolsep8pt
\begin{tabular}{@{}llll@{}}\toprule 

Model & Parameter 1 & Values 1 \\ \toprule 

Clayton-boost &  num\_boost\_round & (0,5000)  \\

Std-boost & num\_boost\_round & (0,5000)  \\

Std-AFT & penalizer & $[0,0.25,0.5,1,2]$  \\

Cox & penalizer & $[0,0.25,0.5,1,2]$\\  \toprule 
\label{tab:cv_sim}
\end{tabular}}
\end{table} 

When comparing the models on the different copula induced data, the Gumbel, Frank and Independent copulas are simulated according Algorithm \ref{alg:gumb}, \ref{alg:frank} and \ref{alg:ind}, respectively. 

\begin{algorithm}
\caption{Simulating from the Gumbel Copula}\label{alg:gumb}
\begin{algorithmic}
\item Input $\theta \geq 1$
\item Draw $V_1 \sim Uniform(0,1)$
\item Draw $V_2 \sim Uniform(0,1)$
\item Draw $Z \sim Levy\_Stable(\frac{1}{\theta},1,(cos(\frac{\pi}{2\theta}))^{\theta},0)$
\item Set $U_1 \gets e^{-(-\frac{log(V_1)}{Z})^{\frac{1}{\theta}}}$
\item Set $U_2 \gets e^{-(-\frac{log(V_2)}{Z})^{\frac{1}{\theta}}}$
\item Return $U_1, U_2$
\end{algorithmic}
\end{algorithm}

\begin{algorithm}
\caption{Simulating from the Frank Copula}\label{alg:frank}
\begin{algorithmic}
\item Input $\theta \neq 0$
\item Draw $V \sim Uniform(0,1)$
\item Draw $U_1 \sim Uniform(0,1)$
\item Set $U_2 \gets -\frac{1}{\theta}log(1+\frac{(1-e^{-\theta})V}{V(e^{-\theta U_1}-1)-e^{-\theta U_1}})$
\item Return $U_1, U_2$
\end{algorithmic}
\end{algorithm}

\begin{algorithm}
\caption{Simulating from the Independent Copula}\label{alg:ind}
\begin{algorithmic}
\item Draw $U_1 \sim Uniform(0,1)$
\item Draw $U_2 \sim Uniform(0,1)$
\item Return $U_1, U_2$
\end{algorithmic}
\end{algorithm}

\subsection{Real data example set-up} \label{sec:appendix_histograms}

The full experiments on the real datasets can be found in the Jupyter Notebooks on this GitHub page: \url{https://github.com/alimid/clayton_boost/tree/main/real_data_examples}.

\begin{figure}
\begin{minipage}{.5\linewidth}
\centering
\subfloat[Z]{\label{hist:gbsg_t}\includegraphics[scale=.45]{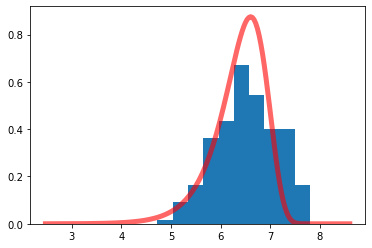}}
\end{minipage}%
\begin{minipage}{.5\linewidth}
\centering
\subfloat[V]{\label{hist:gbsg_u}\includegraphics[scale=.45]{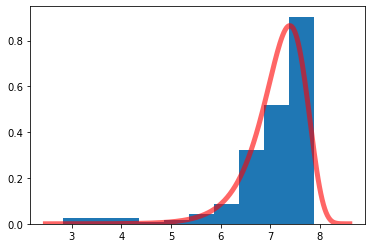}}
\end{minipage}\par\medskip

\caption{Histogram of the logarithm of the event times and censoring times for the GBSG2 training data with the fitted probability distributions.}
\label{fig:hist_gbsg}
\end{figure}

\begin{figure}
\begin{minipage}{.5\linewidth}
\centering
\subfloat[Z]{\label{hist:dd_t}\includegraphics[scale=.45]{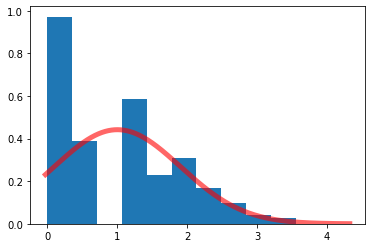}}
\end{minipage}%
\begin{minipage}{.5\linewidth}
\centering
\subfloat[V]{\label{hist:dd_u}\includegraphics[scale=.45]{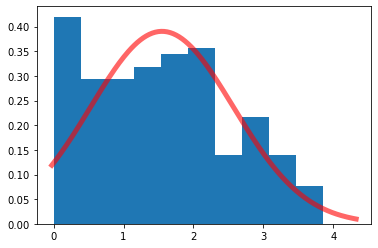}}
\end{minipage}\par\medskip

\caption{Histogram of the logarithm of the event times and censoring times for the DD training data with the fitted probability distributions.}
\label{fig:hidt_dd}
\end{figure}

\begin{figure}
\begin{minipage}{.5\linewidth}
\centering
\subfloat[Z]{\label{hist:fric_t}\includegraphics[scale=.45]{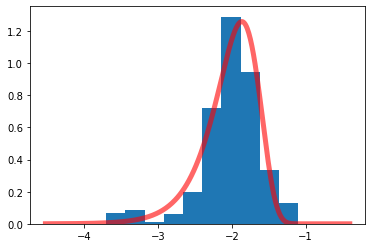}}
\end{minipage}%
\begin{minipage}{.5\linewidth}
\centering
\subfloat[V]{\label{hist:fric_u}\includegraphics[scale=.45]{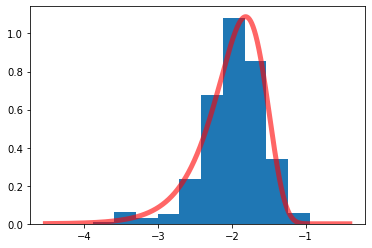}}
\end{minipage}\par\medskip

\caption{Histogram of the logarithm of the event times and censoring times for the FRICTION training data with the fitted probability distributions.}
\label{fig:hist_fric}
\end{figure}

For all the real datasets, the baseline distribution for the AFT models were set empirically by looking at the histograms of the event times and the censoring times. The probability distributions were first fitted automatically by using the python package scipy.stats, and then adjusted according to a visual inspection. The histograms, and the fitted probability distributions, are shown in Figure \ref{fig:hist_gbsg}, \ref{fig:hidt_dd} and \ref{fig:hist_fric}.

\begin{table}
\caption{Chosen baseline distributions for the datasets.}
{\tabcolsep8pt
\begin{tabular}{@{}lllll@{}}\toprule 

Dataset & Z & V & $\sigma_Z$ & $\sigma_V$ \\ \toprule 

GSBG2 & Gumbel & Gumbel & 0.539 & 0.545  \\

DD & Normal & Normal & 0.903 & 1.022  \\

FRICTION & Gumbel & Gumbel & 0.375 & 0.437 \\ \toprule 
\label{tab:ap_data}
\end{tabular}}
\end{table}

The standard deviation of the fitted probability distributions are shown in Table \ref{tab:ap_data}. The Clayton-boost model is affected by the choice of $Z$, $V$ and the values $\sigma_Z$ and $\sigma_V$, the std-boost model is affected by the choice of $Z$ and the value of $\sigma_Z$, the AFT model is affected only by the choice of $Z$ (as it calculates its own standard deviation), while the Cox model is unaffected by all. 

To create a fair comparison, all the four models were tuned with a cross validation grid search with two tuning parameters. The grid search is optimized on maximizing the c-index, and the settings is shown in Table \ref{tab:cv}.

\begin{table}
\caption{Parameter tuning for the different algorithms for the real-world datasets.}
{\tabcolsep8pt
\begin{tabular}{@{}lllll@{}}\toprule 

Model & Parameter 1 & Values 1 & Parameter 2 & Values 2 \\ \toprule 

Clayton-boost & $\theta$ & $[1.1,1.3,1.5,1.8,2.0]$ & num\_boost\_round & (0,5000)  \\

Std-boost & learning\_rate & $[0.2,0.4,0.6,0.8,1]$ & num\_boost\_round & (0,5000)  \\

Std-AFT & penalizer & $[0,0.25,0.5,1,2]$ & l1\_ratio & $[0,0.25,0.5,0.75,1]$ \\

Cox & penalizer & $[0,0.25,0.5,1,2]$ & l1\_ratio & $[0,0.25,0.5,0.75,1]$\\  \toprule 
\label{tab:cv}
\end{tabular}}
\end{table} 

\FloatBarrier

\subsection{Calibration plots} \label{apx:calibration_plots}

\begin{figure}
\begin{minipage}{.33\linewidth}
\centering
\subfloat[$\theta=1e^{-10}$]{\label{ap_theta:a}\includegraphics[scale=.35]{figures/fig3.png}}
\end{minipage}%
\begin{minipage}{.33\linewidth}
\centering
\subfloat[$\theta=1$]{\label{ap_theta:b}\includegraphics[scale=.35]{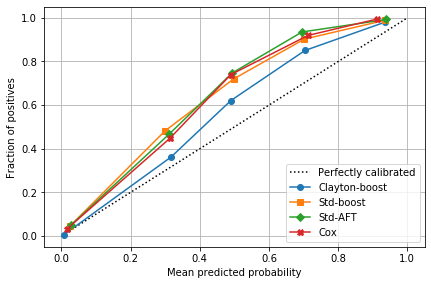}}
\end{minipage}
\begin{minipage}{.33\linewidth}
\centering
\subfloat[$\theta=2$]{\label{ap_theta:c}\includegraphics[scale=.35]{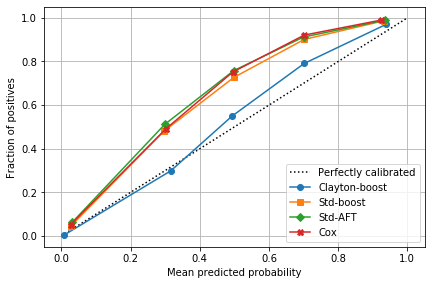}}
\end{minipage}\par\medskip

\begin{minipage}{.33\linewidth}
\centering
\subfloat[$3$]{\label{ap_theta:d}\includegraphics[scale=.35]{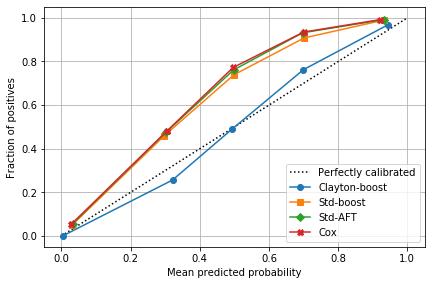}}
\end{minipage}%
\begin{minipage}{.33\linewidth}
\centering
\subfloat[$\theta=4$]{\label{ap_theta:e}\includegraphics[scale=.35]{figures/fig4.png}}
\end{minipage}
\begin{minipage}{.33\linewidth}
\centering
\subfloat[$\theta=5$]{\label{ap_theta:f}\includegraphics[scale=.35]{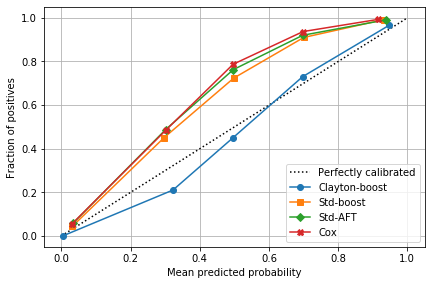}}
\end{minipage}\par\medskip

\begin{minipage}{.33\linewidth}
\centering
\subfloat[$\theta=6$]{\label{ap_theta:g}\includegraphics[scale=.35]{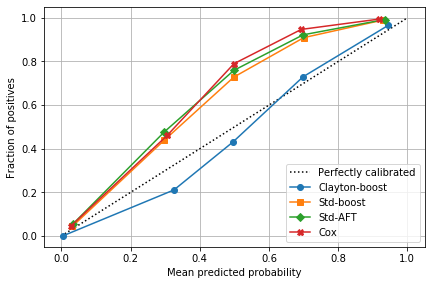}}
\end{minipage}%
\begin{minipage}{.33\linewidth}
\centering
\subfloat[$\theta=7$]{\label{ap_theta:h}\includegraphics[scale=.35]{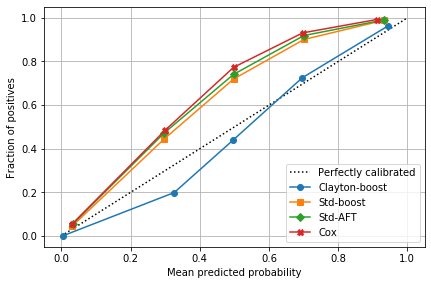}}
\end{minipage}
\begin{minipage}{.33\linewidth}
\centering
\subfloat[$\theta=8$]{\label{ap_theta:i}\includegraphics[scale=.35]{figures/fig5.png}}
\end{minipage}\par\medskip

\caption{Calibration plots for the models under data simulated from different $\theta$ with approx. 50\% censoring}
\label{fig:app_theta}
\end{figure}

\begin{figure}
\begin{minipage}{.33\linewidth}
\centering
\subfloat[$10\%$]{\label{ap_per:a}\includegraphics[scale=.35]{figures/fig8.png}}
\end{minipage}%
\begin{minipage}{.33\linewidth}
\centering
\subfloat[$20\%$]{\label{ap_per:b}\includegraphics[scale=.35]{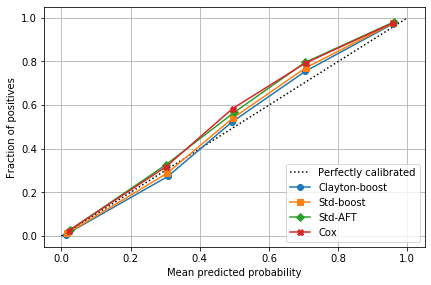}}
\end{minipage}
\begin{minipage}{.33\linewidth}
\centering
\subfloat[$30\%$]{\label{ap_per:c}\includegraphics[scale=.35]{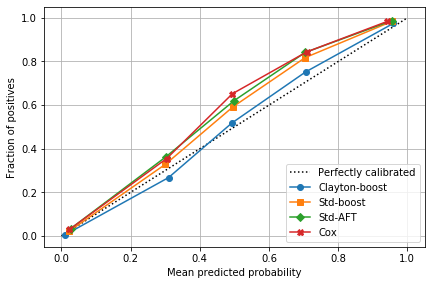}}
\end{minipage}\par\medskip

\begin{minipage}{.33\linewidth}
\centering
\subfloat[$40\%$]{\label{ap_per:d}\includegraphics[scale=.35]{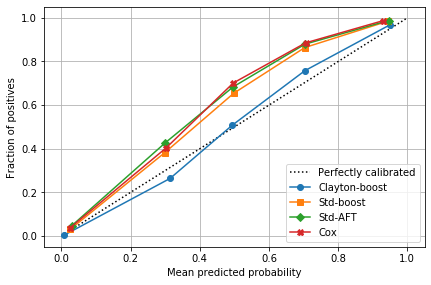}}
\end{minipage}%
\begin{minipage}{.33\linewidth}
\centering
\subfloat[$50\%$]{\label{ap_per:e}\includegraphics[scale=.35]{figures/fig9.png}}
\end{minipage}
\begin{minipage}{.33\linewidth}
\centering
\subfloat[$60\%$]{\label{ap_per:f}\includegraphics[scale=.35]{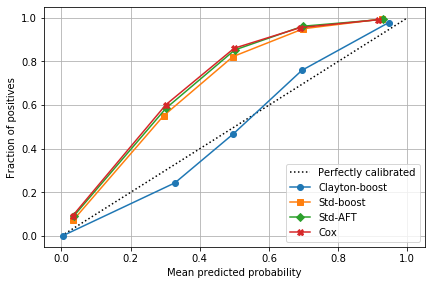}}
\end{minipage}\par\medskip

\begin{minipage}{.33\linewidth}
\centering
\subfloat[$70\%$]{\label{ap_per:g}\includegraphics[scale=.35]{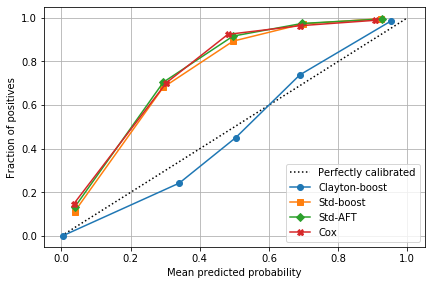}}
\end{minipage}%
\begin{minipage}{.33\linewidth}
\centering
\subfloat[$80\%$]{\label{ap_per:h}\includegraphics[scale=.35]{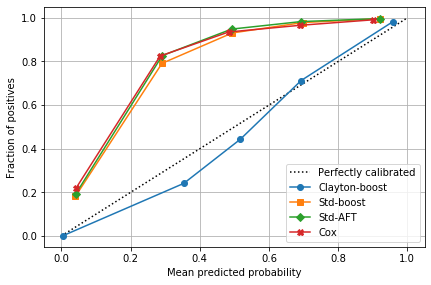}}
\end{minipage}
\begin{minipage}{.33\linewidth}
\centering
\subfloat[$90\%$]{\label{ap_per:i}\includegraphics[scale=.35]{figures/fig10.png}}
\end{minipage}\par\medskip

\caption{Calibration plots for the models under data simulated from different percentage censoring with $\theta=3$}
\label{apx:cal_theta}
\end{figure}


\end{document}